\documentclass[aps,prl,preprint,nofootinbib,nobibnotes,superscriptaddress,notitlepage,showkeys]{revtex4}

\usepackage{amsmath}
\usepackage{amssymb}
\usepackage{graphicx}
\usepackage{epstopdf}
\usepackage{color}
\usepackage[normalem]{ulem}
\usepackage{soul}
\usepackage[FIGTOPCAP]{subfigure}
\usepackage{hyperref}

\def\nat{Nature\ }
\def\aap{Astron.\ Astrophys.\ }
\def\apj{Astrophys.\ J.\ }
\def\apjl{Astrophys.\ J.\ Lett.\ }

\def\app{Astropart.\ Phys.\ }
\def\mnras{Mon.\ Not.\ Roy.\ Astron.\ Soc.\ }
\def\physrep{Phys.\ Rept.\ }
\def\prd{Phys.\ Rev.\ D\ }
\def\prl{Phys.\ Rev.\ Lett.\ }
\def\jcap{J.\ Cos.\ Astropart.\ Phys.\ }

\def\raa{Res.\ Astron.\ Astrophys.\ }
\def\nima{Nucl.\ Instrum.\ Meth.\ A.\ }

\begin{document}

\title{Detection of spectral hardenings in cosmic-ray boron-to-carbon and boron-to-oxygen flux ratios with DAMPE
}

\author{
DAMPE Collaboration$^{\ast}$
}

\footnotetext[1]{Members of DAMPE Collaboration are listed at the end of this paper. \\Email: dampe@pmo.ac.cn}

\begin{abstract}
Boron nuclei in cosmic rays (CRs) are believed to be mainly produced by the fragmentation of heavier nuclei, such as carbon and oxygen, via collisions with the interstellar matter. Therefore, the boron-to-carbon flux ratio (B/C) and the boron-to-oxygen flux ratio (B/O) are very essential probes of the CR propagation. The energy dependence of the B/C ratio from previous balloon-borne and space-based experiments can be well described by a single power-law up to about 1 TeV/n within uncertainties.
This work reports direct measurements of B/C and B/O in the energy range from 10 GeV/n to 5.6 TeV/n with 6 years of data collected by the Dark Matter Particle Explorer, with high statistics and well controlled systematic uncertainties. 
The energy dependence of both the B/C and B/O ratios can be well fitted by a broken power-law model rather than a single power-law model, suggesting the existence in both flux ratios of a spectral hardening at about 100 GeV/n. The significance of the break is about $5.6\sigma$ and $6.9\sigma$ for the GEANT4 simulation, and $4.4\sigma$ and $6.9\sigma$ for the alternative FLUKA simulation, for B/C and B/O, respectively.
These results deviate from the predictions of conventional turbulence theories of the interstellar medium, which point toward a change of turbulence properties of the interstellar medium (ISM) at different scales or novel propagation effects of CRs, and should be properly incorporated in the indirect detection of dark matter via anti-matter particles.
\end{abstract}
\keywords{DAMPE; commic-ray; CR propagation; B/C ratio; B/O ratio}

\maketitle

\section{Introduction}

Galactic CRs are energetic particles travelling through the interstellar space. They are messengers of the violent evolution of stars or stellar systems in extreme environments. CRs are typically divided into two classes, the primary and secondary families. 
Primary CRs are accelerated at astrophysical sources such as supernova remnants, while secondaries are produced from the interactions of the primaries with the interstellar medium (ISM) during the propagation \cite{2007ARNPS..57..285S,2020PhR...872....1B}. The spectrum of accelerated particles at the source is expected to follow a power-law form ${\cal R}^{-p}$ according to the Fermi acceleration mechanism \cite{1949PhRv...75.1169F}, where ${\cal R}$ is the rigidity and $p$ is the power-law index. After the diffusive propagation in the ISM, the spectrum of primary CRs would soften to be $\propto {\cal R}^{-(p+\delta)}$, where $\delta$ is the slope of the rigidity-dependence of the diffusion coefficient. The parameter $\delta$ depends on the power spectrum of the turbulence of the ISM, with typical values of 1/3 for the Kolmogorov theory of interstellar turbulence \cite{1941DoSSR..30..301K} or 1/2 for the Kraichnan theory \cite{1965PhFl....8.1385K}. 
The spectrum of secondary CRs generated by the interaction of primary particles with the ISM is expected to be even softer, $\propto {\cal R}^{-(p+2\delta)}$. The flux ratio of the secondary-to-primary CRs is then $\propto {\cal R}^{-\delta}$, which sensitively depends on the propagation procedure. Precise measurements of the secondary-to-primary flux ratios are thus crucial to reliably constrain the propagation process of CRs \cite{2007ARNPS..57..285S,2020PhR...872....1B}. 

Lithium, beryllium, and boron nuclei in CRs are dominantly produced by the fragmentation of heavier nuclei, since their primary abundances from stellar nucleosynthesis are many orders of magnitude lower than those of protons, helium, carbon, and oxygen. Among all the secondary-to-primary ratios, the B/C ratio is the most extensively measured. The B/O is in principle more directly related to the propagation procedure of CRs than B/C, due to that there is a small amount of secondary contribution for the carbon nuclei. Thanks to the contributions from worldwide experiments, the B/C ratio has been measured up to a few TeV/n \cite{1990A&A...233...96E,1990ApJ...349..625S,2010ApJ...724..329A,2008ICRC....2....3P,2008APh....30..133A,2011ApJ...742...14O,2014ApJ...791...93A,AMS:2016brs,2019AdSpR..64.2559G,2018PhRvL.120b1101A,2021PhR...894....1A}, although the uncertainties are relatively large for kinetic energies above 500 GeV/n. A power-law decline form, $\propto {\cal R}^{-1/3}$, can well fit the rigidity (energy) dependence of the B/C ratio, in agreement with the prediction of the Kolmogorov turbulence \cite{AMS:2016brs}. 
Nevertheless, evidence of breaks of the secondary-to-primary flux ratios was shown by the AMS-02 measurements \cite{2018PhRvL.120b1101A,2021PhR...894....1A}, though the break is not significant for individual B/C or B/O ratio. Improved measurements of the secondary-to-primary ratios, especially towards higher energies, are highly necessary to 
further understand the propagation of CRs and the properties of the interstellar medium.

\section{Results}

In this work, we report the direct measurements of B/C and B/O with the DArk Matter Particle Explorer (DAMPE; also known as ``Wukong''), a satellite-borne detector for high energy cosmic-ray and $\gamma$-ray observations \cite{2017APh....95....6C}. The DAMPE payload consists of a Plastic Scintillator Detector (PSD) for the charge measurement, a Silicon Tungsten tracKer-converter (STK) for the trajectory reconstruction, a bismuth germanium oxide (BGO) imaging calorimeter for the energy measurement and electron-hadron discrimination, and a NeUtron Detector (NUD) to enhance electron-hadron separation \cite{2017APh....95....6C,2017Natur.552...63D}. With its relatively large geometric factor, good charge \cite{2019APh...105...31D} and energy resolution \cite{2017APh....95....6C}, DAMPE is expected to extend the precise measurements of individual spectra of high-abundance CR species from protons to Iron nuclei up to a few hundreds of TeV energies \cite{2019SciA....5.3793A,2021PhRvL.126t1102A}. 
The DAMPE satellite was launched into a 500-km Sun-synchronous orbit on 17 December 2015, and has operated stably in space since then, as illustrated by the on-orbit calibration \cite{2019APh...106...18A}.

\begin{figure}[!ht]
\begin{center}
\includegraphics[width=0.48\columnwidth]{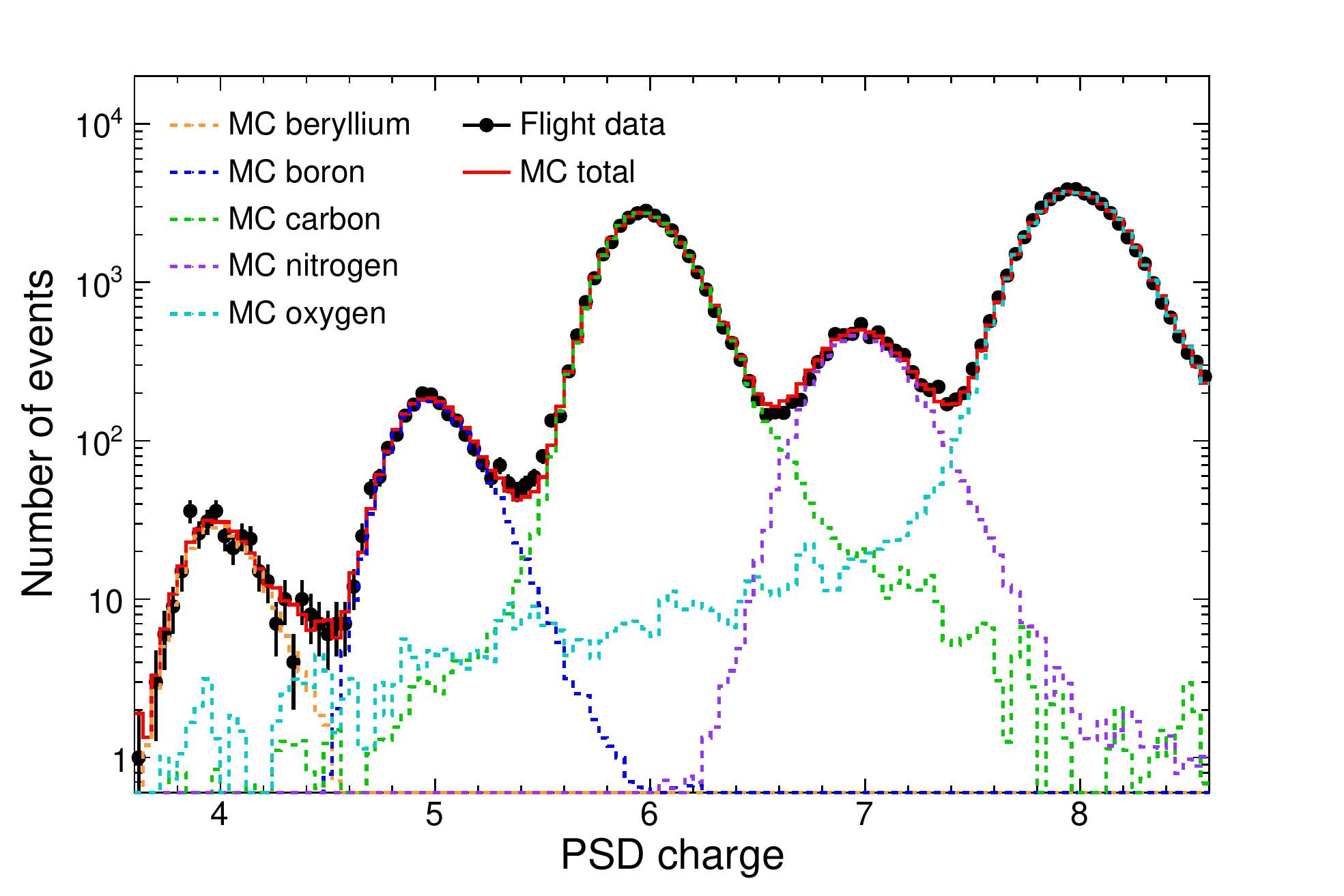}
\put(-225,145){\color{black}{\bf (a)}}
\includegraphics[width=0.48\columnwidth]{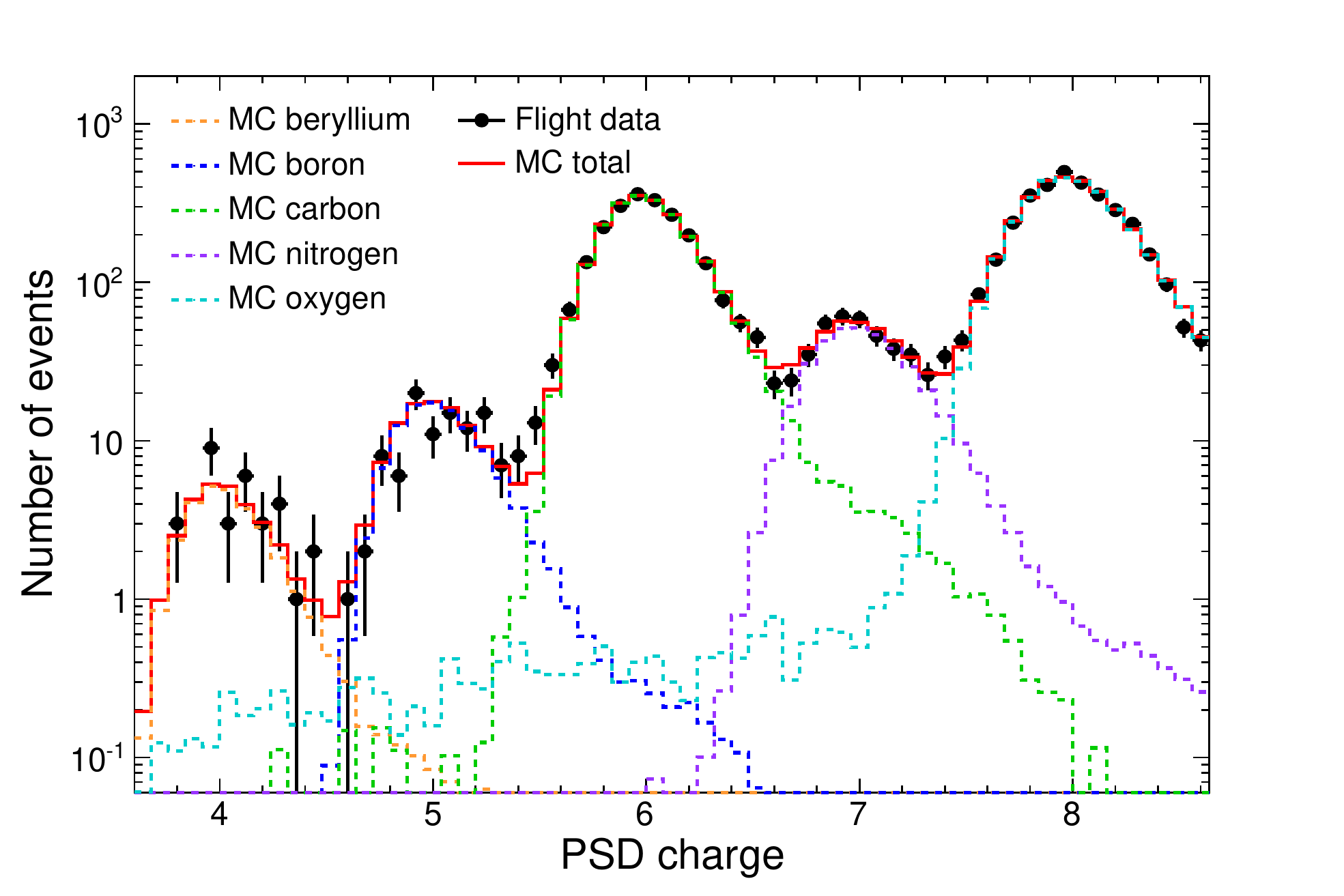}
\put(-225,145){\color{black}{\bf (b)}}
\end{center}
\caption{
{\bf The charge distributions measured by PSD for particles with $Z=4-8$ and deposited energies in the calorimeter of 630 GeV to 2 TeV (a), and 3.16 TeV to 10 TeV (b).} The flight data are shown by black dots. Dashed lines with different colors show the best-fit MC simulated samples of beryllium, boron, carbon, nitrogen, and oxygen nuclei. The sum of MC samples is shown by the red line. 
}
\label{fig-1}
\end{figure}

The analysis presented in this work is based on the data recorded in the first 6 years of DAMPE's operation, from January 1, 2016 to December 31, 2021. The live time fraction is about 75.85\% after excluding the instrument dead time, the time for the on-orbit calibration, the time in the South Atlantic Anomaly (SAA) region, and the period between September 9, 2017 and September 13, 2017 during which a big solar flare affected the status of the detector \cite{DAMPE:2021qet}. The boron, carbon, and oxygen nuclei are efficiently identified based on the PSD charge measurement.  Fig.~\ref{fig-1} illustrate the reconstructed PSD charge distributions for events with $Z=4-8$ and deposited energies in the calorimeter of 630 GeV to 2 TeV, and 3.16 TeV to 10 TeV. The Monte Carlo (MC) simulations for nuclei from beryllium to oxygen, generated with {\tt GEANT} v4.10.05 \cite{GEANT4:2003zbu}, are shown by dashed lines to illustrate a best-fit to the flight data. Here, we suppress lighter nuclei $(Z<4)$ using a STK charge selection (see Supplementary Material for details). Residual nuclei lighter than beryllium are too low to be shown in these plots.

\begin{figure}[!ht]
\begin{center}
\includegraphics[width=0.48\columnwidth]{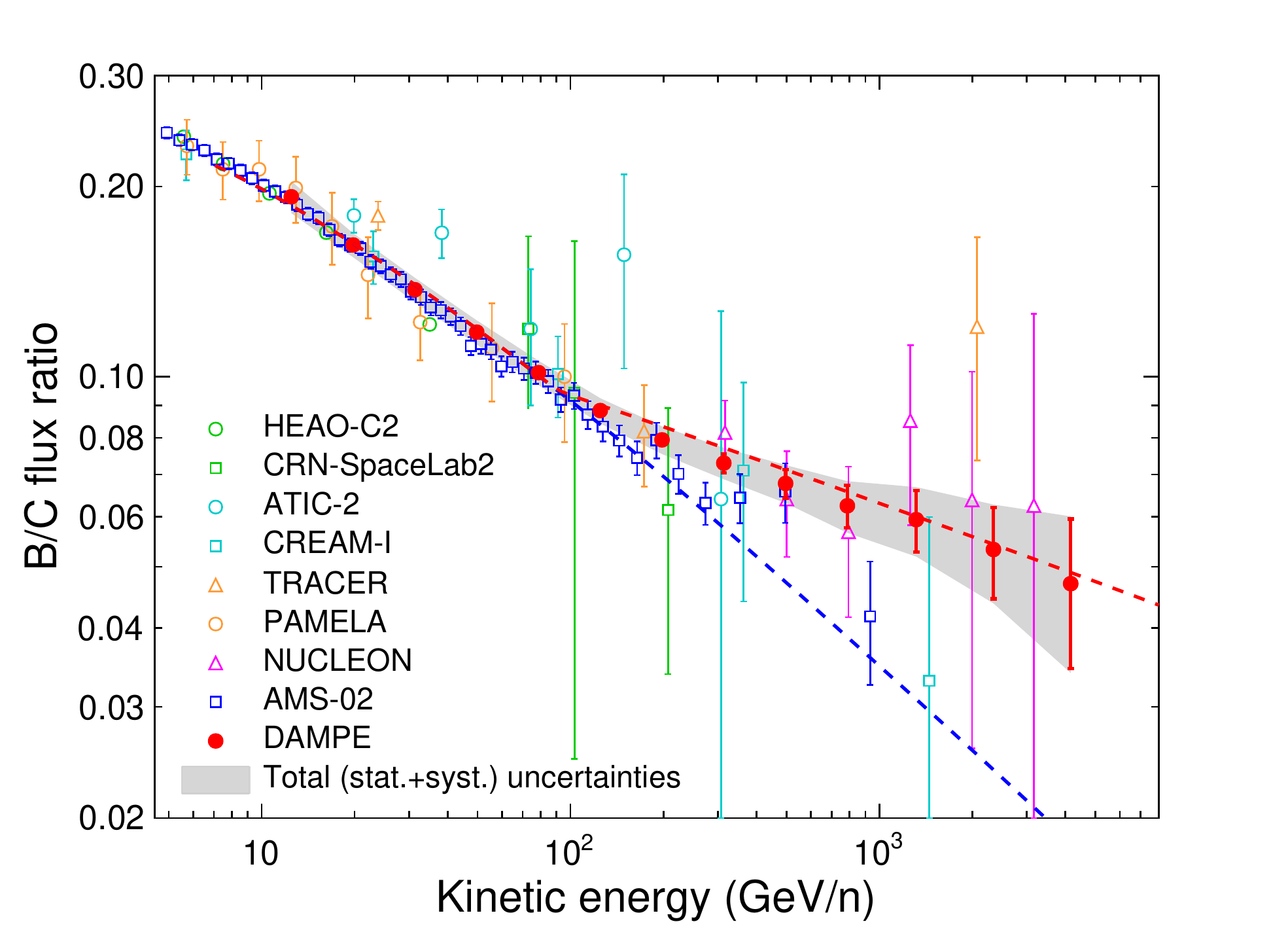}
\put(-225,155){\color{black}{\bf (a)}}
\includegraphics[width=0.48\columnwidth]{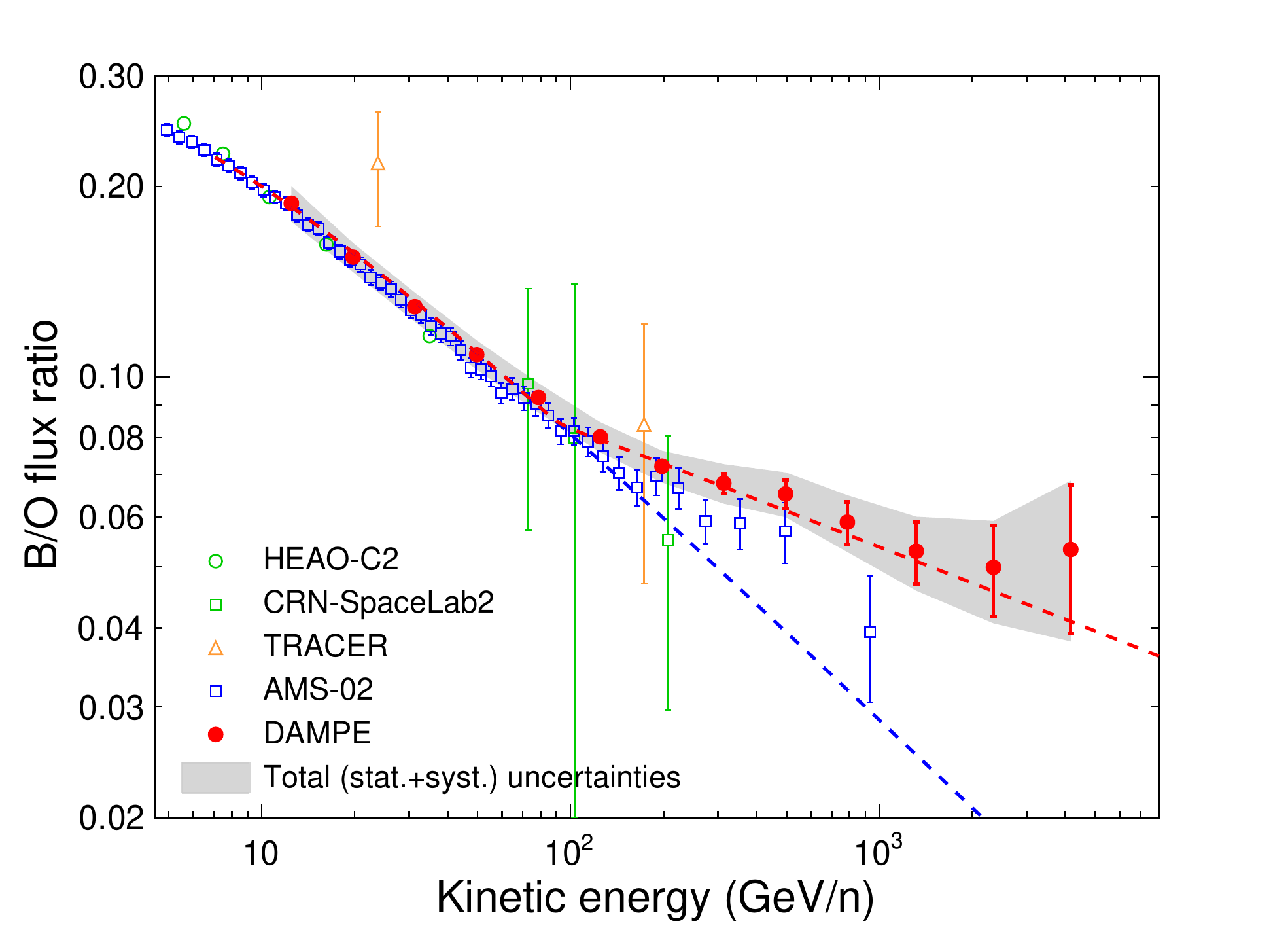}
\put(-225,155){\color{black}{\bf (b)}}
\end{center}
\caption{
{\bf Boron-to-carbon (a) and boron-to-oxygen (b) flux ratios as functions of kinetic energy per nucleon.} DAMPE measurements are shown by red filled dots, with error bars and shaded bands representing the statistical and total uncertainties, respectively. The total uncertainties are the sum in quadrature of the statistical and systematic ones.
The blue dashed lines show the fitting results for a GALPROP model with single power-law rigidity dependence of the diffusion coefficient, and the red dashed lines are the results with a hardening of the diffusion coefficient at $200$ GV. In panel (a), other direct measurements by HEAO3 \cite{1990A&A...233...96E} (green circles), CRN \cite{1990ApJ...349..625S} (green squares), ATIC-2 \cite{2008ICRC....2....3P} (cyan circles), CREAM-I \cite{2008APh....30..133A} (cyan squares), TRACER \cite{2011ApJ...742...14O} (orange triangles), PAMELA \cite{2014ApJ...791...93A} (orange circles),  NUCLEON-KLEM \cite{2019AdSpR..64.2559G} (magenta triangles) and AMS-02 \cite{2021PhR...894....1A} (blue squares) are shown for comparison. In panel (b), the measurements of B/O by HEAO3 \cite{1990A&A...233...96E} (green circles), CRN \cite{1990ApJ...349..625S} (green squares), TRACER \cite{2011ApJ...742...14O} (orange triangles) and AMS-02 \cite{2021PhR...894....1A} (blue squares) are shown. 
For the AMS-02 results \cite{2021PhR...894....1A}, we convert the ratios from rigidity to kinetic energy per nucleon assuming an atomic mass number of 10.7 for boron, 12.0 for carbon, 16.0 for oxygen, and a power-law spectrum of carbon (oxygen) with an index of $-2.6$.
The error bars of TRACER, CREAM-I, PAMELA, and AMS-02 data include both statistical and systematic uncertainties added in quadrature. For HEAO3, CRN, ATIC-2, and NUCLEON data only the statistical uncertainties are shown.}
\label{fig-2}
\end{figure}

The boron, carbon, and oxygen candidates are selected with energy-independent charges of [4.7, 5.3], [5.6, 6.4], and [7.6, 8.5], respectively. The total contamination of the boron sample is found to be $\sim1\%$ for deposited energies around 100 GeV and $\sim4.5\%$ around 50 TeV, while the contamination of the carbon and oxygen sample is $<0.6\%$ and $<1.6\%$ respectively, over the entire energy range. In Fig.~\ref{fig-1}, the distribution of MC oxygen shows a more prominent tail on the lower charge side compared with those from other nuclei, which is primarily due to their different fragmentation cross sections with the materials above or in the PSD. As a result, the contamination to boron from oxygen is larger than that from carbon. Similar distributions are also shown for the {\tt FLUKA} \cite{Bohlen:2014buj} simulations, although the inelastic interactions of {\tt FLUKA} and {\tt GEANT4} are different. 

The selection efficiency and the energy response of the calorimeter are obtained with MC simulations, and validated from the flight data and the test beam data. Applying an unfolding procedure \cite{1995NIMPA.362..487D}, we derive the B/C and B/O ratios in the energy range from 10 GeV/n to 5.6 TeV/n, as shown in  Fig.~\ref{fig-2} and tabulated in Table \ref{tab:Table1}. The atomic mass numbers are assumed to be 10.7 (see Ref.~\cite{AMS:2016brs}), 12, and 16 for boron, carbon, and oxygen, respectively. Compared with previous measurements by HEAO3 \cite{1990A&A...233...96E}, CRN \cite{1990ApJ...349..625S}, ATIC-2 \cite{2008ICRC....2....3P}, CREAM-I \cite{2008APh....30..133A}, TRACER \cite{2011ApJ...742...14O}, PAMELA \cite{2014ApJ...791...93A}, AMS-02 \cite{2021PhR...894....1A}, and NUCLEON \cite{2019AdSpR..64.2559G}, the DAMPE measurements are well consistent with them at low energies ($E_k\lesssim500$ GeV/n) and improve the precision significantly at high energies. Particularly, the DAMPE results provide the first precise measurements of the B/C and B/O ratios above 1 TeV/n. 

\begin{table}[ht]
\begin{center}
\caption{{\bf Boron-to-carbon and boron-to-oxygen flux ratios measured with DAMPE, together with 1$\sigma$ statistical and systematic uncertainties.}
}
\begin{tabular}{ccccc}
\hline\hline
$\langle E \rangle$ & $E_{\rm min}$ & $E_{\rm max}$ & B/C & B/O\\
(GeV/n) & (GeV/n) & (GeV/n) & ${\rm ratio} \pm \sigma_{\rm stat} \pm \sigma_{\rm sys}$ & 
${\rm ratio} \pm \sigma_{\rm stat} \pm \sigma_{\rm sys}$ \\
\hline
12.5 & 10.0 & 15.8 & $0.1926 \pm 0.0017 \pm 0.0111$ & $0.1882 \pm 0.0025 \pm 0.0119$ \\
19.8 & 15.8 & 25.1 & $0.1616 \pm 0.0007 \pm 0.0070$ & $0.1546 \pm 0.0008 \pm 0.0081$ \\
31.3 & 25.1 & 39.8 & $0.1373 \pm 0.0006 \pm 0.0061$ & $0.1290 \pm 0.0007 \pm 0.0068$ \\
49.7 & 39.8 & 63.1 & $0.1176 \pm 0.0007 \pm 0.0051$ & $0.1084 \pm 0.0008 \pm 0.0057$ \\
78.7 & 63.1 & 100 & $0.1015 \pm 0.0010 \pm 0.0044$ & $0.0927 \pm 0.0010 \pm 0.0049$ \\ 
125 & 100 & 158 & $0.0884 \pm 0.0013 \pm 0.0038$ & $0.0803 \pm 0.0012 \pm 0.0042$ \\
198 & 158 & 251 & $0.0794 \pm 0.0018 \pm 0.0036$ & $0.0722 \pm 0.0017 \pm 0.0038$ \\ 
313 & 251 & 398 & $0.0730 \pm 0.0025 \pm 0.0033$ & $0.0678 \pm 0.0024 \pm 0.0043$ \\
497 & 398 & 631 & $0.0678 \pm 0.0035 \pm 0.0031$ & $0.0652 \pm 0.0034 \pm 0.0041$ \\
787 & 631 & 1000 & $0.0624 \pm 0.0048 \pm 0.0034$ & $0.0588 \pm 0.0045 \pm 0.0041$ \\
1315 & 1000 & 1778 & $0.0594 \pm 0.0067 \pm 0.0034$ & $0.0529 \pm 0.0059 \pm 0.0039$ \\
2339 & 1778 & 3162 & $0.0532 \pm 0.0088 \pm 0.0036$ & $0.0499 \pm 0.0083 \pm 0.0041$ \\
4160 & 3162 & 5623 & $0.0470 \pm 0.0125 \pm 0.0038$ & $0.0532 \pm 0.0141 \pm 0.0055$\\
\hline\hline
\end{tabular}
\label{tab:Table1}
\end{center}
\end{table}

\section{Discussion and conclusion}

Fits to the DAMPE measurements show that both the energy dependence of B/C and B/O deviate 
from single power-law (PL) forms in the measured energy range. A broken power-law (BPL) model fit 
yields to a $\chi^2=6.61$ for 5 degrees of freedom (dof) while the PL fit yields to a $\chi^{2}=42.35$ 
for 7 dof for B/C, for the {\tt GEANT4} simulation. Similarly, for B/O, the BPL fit gives
$\chi^2/{\rm dof}=5.51/5$ while the PL fit yields $\chi^2/{\rm dof}=57.81/7$. Therefore, the DAMPE 
data favor a spectral break of B/C (B/O) with a significance of $5.6\sigma$ ($6.9\sigma$) through 
comparing the $\Delta\chi^2$ values. The fits to the results with the {\tt FLUKA} simulation
give a significance of $4.4\sigma$ ($6.9\sigma$) for the B/C (B/O) ratio.
The break energy is found to be $98.9^{+8.9+10.0}_{-8.8-0.0}$ 
($99.5^{+7.4+7.7}_{-7.1-0.0}$) GeV/n, and the spectral indices below/above $E_b$ are 
$(\gamma_1,\gamma_2)=(0.356^{+0.008+0.000}_{-0.008-0.017}$, $0.201^{+0.024+0.008}_{-0.024-0.000})$ 
and $(\gamma_1,\gamma_2)=(0.394^{+0.010+0.000}_{-0.010-0.026}, 0.187^{+0.024+0.000}_{-0.024-0.019})$ 
for B/C and B/O, respectively (see Supplementary Material for details). Here, the first error 
comes from the fitting and the second error comes from the comparison with the alternative 
analysis based on the {\tt FLUKA} simulation. We find that the break energies and the high-energy 
spectral indices of B/C and B/O are consistent with each other, while the low-energy spectral 
index of B/C is slightly harder than that of B/O. The difference may come from the fact that 
the carbon spectrum is softer than the oxygen spectrum below $\sim100$ GeV/n as revealed by 
AMS-02 \cite{2021PhR...894....1A} and CALET \cite{Adriani:2020wyg}, which may be due to a small 
secondary contribution of carbon from oxygen and heavier nuclei. The corresponding spectral 
index changes are found to be  $\Delta\gamma=0.155^{+0.026+0.000}_{-0.026-0.026}$
($\Delta\gamma=0.207^{+0.027+0.000}_{-0.028-0.007}$) for B/C (B/O).

The DAMPE results have far-reaching implications on the propagation of Galactic CRs. The slope
parameter $\delta$ of the diffusion coefficient is predicted to be either 1/3 or 1/2 in the
conventional turbulence theories \cite{1941DoSSR..30..301K,1965PhFl....8.1385K}. The detection of
spectral hardenings in the B/C and B/O ratios by DAMPE thus challenges these conventional scenarios. 
To introduce a spectral break of the diffusion coefficient may be the simplest solution to account 
for the observations\cite{Genolini:2017dfb}. We have illustrated in Fig.~\ref{fig-2} that the 
fitting to the pre-DAMPE data with a single power-law form of the diffusion coefficient, 
$D(R)\propto R^{\delta}$ with $\delta=0.477$ \cite{2020JCAP...11..027Y}, using the GALPROP
model \cite{1998ApJ...509..212S} assuming the convective transportation of CRs, deviates clearly 
from the DAMPE high-energy measurements (see the blue dashed lines). If we add a spectral break 
at $R_{\rm br}=200$ GV, with a high-energy slope $\delta_h=0.2$, the model prediction matches well 
with the measurements as shown by the red dashed lines. Intriguingly, the inferred $\delta=0.477$ 
at rigidities of $\leq 200$ GV is very close to the prediction of the Kraichnan theory of 
turbulence \cite{1965PhFl....8.1385K}. At higher rigidities, the rigidity dependence of $R^{-0.2}$ 
is harder than that expected by the Kolmogorov theory of turbulence \cite{1941DoSSR..30..301K}. 
This deviation may be relieved if a small amount of secondary particles were generated at the 
sources (i.e., they experience the same propagation process and thus give rise to a constant, 
although small, ratio). Our findings may thus imply the change of turbulence properties of the 
ISM at different scales, e.g., from the magnetized turbulence (Kraichnan type) at smaller scales 
to isotropic, stationary hydrodynamic turbulence (Kolmogorov type) at larger scales.

Alternatively, more complicated propagation or acceleration effects of CRs may also result in 
hardenings of the secondary-to-primary ratios. These models include, but are not limited to, 
the nested leaky box propagation model with different energy-dependence of the residence time 
in the ISM and the cocoon regions surrounding the sources \cite{2016ApJ...827..119C}, 
the production and acceleration of secondary particles at sources \cite{2021PhRvD.104j3029M}, 
the re-acceleration of CRs by random magnetohydrodynamic waves during the
propagation \cite{2020JCAP...11..027Y} or by a local shock \cite{2022ApJ...933...78M}, the self-generation of turbulence by CRs \cite{2012PhRvL.109f1101B}, the spatially-dependent diffusion of
particles \cite{2012ApJ...752L..13T}, or possibly, a mixture of some of them \cite{Bresci:2019aww}.

In addition to the CR propagation studies, a significant spectral hardening of B/C (B/O) 
should be properly addressed in the search of dark matter annihilation or decay products with 
the antiparticle CRs, such as positrons and antiprotons, since the predictions of astrophysical
background and the dark matter induced signal should both be affected by the change of the
diffusion process. For instance, the previously claimed excess in the anti-proton 
data \cite{Cui:2016ppb,2017PhRvL.118s1102C} may need a thorough re-examination to critically
address its potential connection with the dark matter annihilation or decay.
Improved measurements of the B/C, B/O, and other secondary-to-primary ratios with higher 
statistics and lower systematics by DAMPE and future direct detection experiments such as
HERD \cite{2014SPIE.9144E..0XZ}, AMS-100 \cite{Schael:2019lvx}, and ALADInO \cite{Adriani:2022} 
are expected to eventually uncover the fundamental problems of the origin and propagation of 
CRs and shed new light on the indirect detection of dark matter particles.

\section{}

{\bf Conflict of interest:} The authors declare that they have no conflict of interest.

{\bf Acknowledgments:} The DAMPE mission was funded by the strategic priority science and technology projects in space science of Chinese Academy of Sciences (CAS).
In China the data analysis is supported by the National Natural Science Foundation of China (No. 11921003, No. 11903084, No. 12003076, No. 12022503 and No. 12220101003), the strategic priority science and technology projects of CAS (No. XDA15051100), the CAS project for Young Scientists in Basic Research (No. YSBR-061), the Youth Innovation Promotion Association of CAS, the Natural Science Foundation of Jiangsu Province (No. BK20201107), and the Program for Innovative Talents and Entrepreneur in Jiangsu. In Europe the activities and data analysis are supported by the Swiss National Science Foundation (SNSF), Switzerland, the National Institute for Nuclear Physics (INFN), Italy, and the European Research Council (ERC) under the European Union’s Horizon 2020 research and innovation programme (No. 851103).



\section{}
\centerline
{\large\bf The DAMPE experiment}
The DAMPE is the first Chinese astronomical satellite, which consists of four sub-detectors, including the plastic scintillator detector, the silicon tracker, the BGO calorimeter and the neutron detector. The main scientific objectives addressed by DAMPE include probing the dark matter via the detection of high-energy electrons/positrons and gamma-rays, understanding the origin, acceleration and propagation of cosmic rays in the Milky Way, and studying the gamma-ray astronomy. The DAMPE mission is funded by the Strategic Priority Science and Technology Projects in Space Science of the Chinese Academy of Sciences. The DAMPE Collaboration consists of 150 members from 3 countries, including physicists, astrophysicists and engineers.

\section{}
\renewcommand{\thefootnote}{\fnsymbol{footnote}}
\centerline
{\large\bf DAMPE collaboration}
\begin
{flushleft}
\noindent
\small
Francesca~Alemanno$^{1,2}$,
Corrado~Altomare$^{3}$,
Qi~An$^{4,5}$,
Philipp~Azzarello$^{6}$,
Felicia~Carla~Tiziana~Barbato$^{1,2}$,
Paolo~Bernardini$^{7,8}$,
Xiao-Jun~Bi$^{9,10}$,
Ming-Sheng~Cai$^{11,12}$,
Elisabetta~Casilli$^{7,8}$,
Enrico~Catanzani$^{13}$
% ...
\bigskip

{\footnotesize \it
$^1$Gran Sasso Science Institute (GSSI), I-67100 L’Aquila, Italy\\
$^2$Istituto Nazionale di Fisica Nucleare (INFN) - Laboratori Nazionali del Gran Sasso, I-67100 Assergi, L’Aquila, Italy\\
$^3$Istituto Nazionale di Fisica Nucleare, Sezione di Bari, I-70126 Bari, Italy\\
$^4$State Key Laboratory of Particle Detection and Electronics, University of Science and Technology of China, Hefei 230026, China\\
$^5$Department of Modern Physics, University of Science and Technology of China, Hefei 230026, China\\
$^6$Department of Nuclear and Particle Physics, University of Geneva, CH-1211, Switzerland\\
$^7$Dipartimento di Matematica e Fisica E. De Giorgi, Universit\`a del Salento, I-73100, Lecce, Italy\\
$^8$Istituto Nazionale di Fisica Nucleare (INFN) - Sezione di Lecce, I-73100, Lecce, Italy\\
$^9$University of Chinese Academy of Sciences, Beijing 100049, China\\
$^{10}$Particle Astrophysics Division, Institute of High Energy Physics, Chinese Academy of Sciences, Beijing 100049, China\\
$^{11}$Key Laboratory of Dark Matter and Space Astronomy, Purple Mountain Observatory, Chinese Academy of Sciences, Nanjing 210023, China\\
$^{12}$School of Astronomy and Space Science, University of Science and Technology of China, Hefei 230026, China\\
$^{13}$Istituto Nazionale di Fisica Nucleare (INFN) - Sezione di Perugia, I-06123 Perugia, Italy\\
$^{14}$Institute of Modern Physics, Chinese Academy of Sciences, Lanzhou 730000, China\\
$^{15}$National Space Science Center, Chinese Academy of Sciences, Beijing 100190, China\\
$^{16}$Dipartimento di Fisica ``M.~Merlin'', dell’Universit\`a e del Politecnico di Bari, I-70126 Bari, Italy\\
$^{17}$Department of Physics and Laboratory for Space Research, the University of Hong Kong, Hong Kong SAR, China\\
\medskip
}
\end{flushleft}

\clearpage

\section*{Supplementary Material}

\subsection*{MC simulations}
Extensive MC simulations are carried out to estimate the instrument response of incident particles in the DAMPE detector. In this work, the {\tt GEANT} toolkit v4.10.05 \cite{GEANT4:2003zbu} with the {\tt FTFP\_BERT} physics list is adopted for the simulations of nuclei. For higher energies we link the {\tt EPOS\_LHC} model by means of a {\tt CRMC-GEANT4} interface \cite{Tykhonov:2019xlb}. 
The energy response of MC simulations is tuned by including the Birks' quenching \cite{Birks:1951, Wei:2020} for the ionization energy deposits in the BGO calorimeter, due to secondary particles with a large charge number and a low kinetic energy. That correction results in a $\sim 3\%$ decrease of energy deposition at incident energy of 100 GeV and $<1\%$ above 1 TeV for nuclei from boron to oxygen.

The simulated events are generated assuming an isotropic source with an $E^{-1.0}$ spectrum. In the analysis, the simulation data are re-weighted to $E^{-2.6}$ and $E^{-3.0}$ spectra, for primary (e.g. carbon and oxygen) and secondary (e.g. boron) nuclei, respectively. For boron nuclei, $^{10}$B and $^{11}$B samples are mixed assuming an isotopic composition of $Y_{\rm B}$=$^{11}$B/($^{11}$B+$^{10}$B)=0.7, according to the AMS-02 low energy measurements \cite{AMS:2016brs} and also the prediction from nuclear fragmentation \cite{Luque:2022aio}.
As an evaluation of the uncertainties from the hadronic interaction model, we also perform simulations with the {\tt FLUKA} 2011.2x package \cite{Bohlen:2014buj}, which uses {\tt DPMJET3} for nucleus-nucleus interaction above 5 GeV/n. The same analysis procedure based on the two simulations are carried out, and the final differences of the B/C and B/O raitos are taken as systematic uncertainties from the hadronic interaction model.

\subsection*{Event selection} 
DAMPE implements four different triggers on orbit \cite{2019RAA....19..123Z}, among which the high-energy (HE) trigger is chosen to select events for the CR spectral analysis. The events with total deposited energy in the BGO calorimeter ($E_{\rm BGO}$) higher than 80 GeV are selected to avoid the geomagnetic rigidity cut-off effect \cite{2015EP&S...67..158T}.
To ensure a good shower containment, the BGO crystal with the maximum energy deposition in each of the first six layer is required not to be at the edge of the calorimeter.

The trajectory of an incident particle is obtained by optimizing the multiple STK tracks reconstructed with the Kalman filter algorithm \cite{DAMPE:2017yae}.
The quality of the track is evaluated by jointly considering the number of hits on the track, the $\chi^2/{\rm dof}$ value of the Kalman filter, the signal consistency of each hit, and the deviation between the track and the shower axis in the calorimeter. In case that several good track candidates are found, the one with the maximum average hit energy is chosen. The selected track is then required to pass the PSD with maximum energy in both $X$ and $Y$ views, and pass through the calorimeter from top to bottom.

The particle charge $Z$ is reconstructed with the ionization energy deposited in both PSD and STK. The STK consists of six planes of two orthogonal layers of single-sided silicon microstrip detectors \cite{Azzarello:2016trx}. We first require the charge value from the hit of the first STK plane along the track to be larger than 4 ($Q_{\rm STK1st}>4$), in order to effectively suppress particles lighter than boron. Then we employ the PSD hits on the selected track to calculate the particle charge.
The PSD is composed of four sub-layers placed in a hodoscopic configuration in $YZ$-view and $XZ$-view \cite{Yu:2017dpa}, which provides, at most, four independent charge measurements. A detailed charge reconstruction algorithm is applied for each hit based on its ionization energy deposition, including the path length correction, the light attenuation correction and the light yield saturation correction \cite{2019APh...105...31D,2019RAA....19...82M}. We eliminate the energy-dependence of the charge measurements, primarily due to back-scattered secondaries whose signals add up to the primary particle's ionization signal, via setting the peaks to corresponding integer charge values. 
The PSD charge hits on the trajectory are further selected by a consistency requirement of $|\Delta Z|<1$ (sub-layer by sub-layer from top to bottom). The final charge value ($Q_{\rm PSD}$) is obtained by averaging the charge measurements from the selected PSD hits, which achieves a good energy-independence as shown in Fig.~\ref{fig-psdQ}. The same procedure is applied to the MC simulations, and the MC charge distributions are shrinked to match the flight data. The boron, carbon and oxygen candidates are selected with energy-independent charge intervals of [4.7, 5.3], [5.6, 6.4] and [7.6, 8.5], respectively. After the charge selection, we have $1.16 \times 10^{5}$ boron, $1.27 \times 10^{6}$ carbon and $2.17 \times 10^{6}$ oxygen candidates with $E_{\rm BGO}>80$ GeV.

\subsection*{Background estimate}

The background comes from the mis-identification of particle charge, primarily due to the fragmentation in PSD. We employ the MC charge distributions as templates to fit the flight data and estimate the background contributions. The contamination fractions from different species of boron, carbon, and oxygen are shown in Fig.~\ref{fig-bkg}.
The background from nuclei heavier than fluorine is neglected in this analysis, as their fluxes are much lower than those of carbon and oxygen.
The contamination of the boron sample is found to be 1\% to 2\% for $E_{\rm BGO}<1$ TeV and $\sim$4.5\% around 50 TeV, while the contamination of the carbon and oxygen sample is less than 0.6\% and 1.6\% respectively, over the entire energy range.
Tuning the MC charge templates, e.g., via charge-dependent shift and shrink results in $<1\%$ change of the background for boron, and $<0.1\%$ change of those for carbon and oxygen, which is neglected in the present analysis.

\subsection*{Energy measurement and spectral unfolding}

The deposited energy $E_{\rm BGO}$ is obtained as the sum of the energy deposit in each crystal of the calorimeter. The small bias of the absolute energy scale, $\sim$1.25\% as estimated by the geomagnetic cutoff of the electron and positron spectra \cite{Zang2018}, barely affects the flux ratio measurement and is not corrected in this analysis. The designed linear region of the energy measurement of a single BGO bar is $\sim 4$ TeV for the dynode-2 readout. For high energy events, typically above 20 TeV, the energy deposit in a single crystal would exceed the readout upper limit, resulting in a saturation of the energy measurement. To correct this effect, we develop a method based on MC simulations to estimate the energy deposit(s) of saturated crystal(s) \cite{Yue:2020hmj}. The linearity of the energy measurement is validated with the electron test beams up to 243 GeV \cite{ZhangZY2016}. 
At even higher energies no test beam is available to directly validate the energy linearity. We carried out a laser test to study the response of the BGO calorimeter and found that the BGO fluorescence response retains a good linearity at volume energy densities higher by a factor of $\sim5$ than that induced by a 10 TeV electromagnetic shower \cite{ZhaoC2022}. 
Also, we investigate the maximum energy deposited in one single BGO bar versus the total deposited energy to characterize the measurements after applying the saturation correction, as shown in Fig.~\ref{fig-maxbarE}. The quenching effect described above is included in the simulation data. Good consistency between the flight data and the MC simulation indicates that no clear nonlinearity of the measurement exists within the interested energy range of this work.

The BGO energy response to nuclei was studied at CERN SPS in 2014-2015 using beams of accelerated ion fragments with $A/Z=2$ and kinetic energies of 40 and 75 GeV/n \cite{Wei:2019rep,Zhang:2020zuz}. The deposited energy distributions for carbon and oxygen nuclei at 75 GeV/n are shown in Fig.~\ref{fig-beam}. The comparison of the energy response between the beam test data and the {\tt GEANT4 FTFP\_BERT} simulations shows a good agreement within the statistical uncertainties of beam test events. 

Due to the energy leakage of hadronic shower in the calorimeter because of its limited thickness ($\sim$ 1.6 nuclear interaction length), the energy resolution for nuclei measurements is not as good as for electrons/photons. Furthermore, the deposited energy fraction shows a decrease trend with the increase of the incident energy. An unfolding procedure is thus necessary to account for the bin-to-bin migration effect. The {\tt observed} number of events, $N_{{\rm obs},i}$, in the $i$-th deposited energy bin is related to the {\tt incident} number of events, $N_{{\rm inc},j}$, in the $j$-th incident energy bin via the response matrix $M$ as
\begin{equation}
N_{{\rm obs},i}(1-\beta_{\rm i})=\sum_{j} M_{ij} N_{{\rm inc},j},
\end{equation}
where $\beta_{i}$ is the background fraction, $M_{ij}$ is the probability that particles in the $j$-th incident energy bin contributing to the $i$-th deposited energy bin. The response matrix is derived using MC simulations after applying the same selection procedure as for the flight data. In this work, we use the Bayesian unfolding approach \cite{1995NIMPA.362..487D} to derive the {\tt incident} numbers of events. The uncertainty of the energy response matrix, mainly due to the uncertainty of the hadronic interaction model, is estimated through a comparison between different MC simulations, i.e. {\tt GEANT4} and {\tt FLUKA}, and is included in the systematic uncertainties. 

\subsection*{Flux ratio calculation}

In order to obtain the flux ratio as a function of the kinetic energy per nucleon ($E_{k}$), the atomic mass numbers are averaged by assuming an isotope composition from AMS-02
measurements \cite{AMS:2016brs} for boron, pure $^{12}$C for carbon and pure 
$^{16}$O for oxygen. The flux ratio of B/C (B/O) in the $i$-th $E_{k}$ bin (we choose different binnings among B, C, O during the unfolding process to make sure that they have the same $E_k$ binning) is given by 
\begin{equation}
R_{i}= \frac{\Phi^{\rm B}_{i}}{\Phi^{\rm C(O)}_{i}}=\frac{N^{\rm B}_{i}}{N^{\rm C(O)}_{i}} \left ( \frac{\varepsilon^{\rm B}_{i}}{\varepsilon^{\rm C(O)}_{i}} \right )^{-1},
\end{equation}
where $N^{\rm B}_{i}$ and $N^{\rm C(O)}_{i}$ are the unfolded numbers of boron and carbon (oxygen) nuclei, $\varepsilon^{\rm B}_{i}$ and $\varepsilon^{\rm C(O)}_{i}$ are the total selection efficiencies derived from MC simulations. The efficiencies are also validated with the flight data, with deviations being treated as systematic uncertainties. 

\subsection*{Uncertainty analysis}

The statistical uncertainties refer to the Poisson fluctuations of the measured number of events in each deposited energy bin. To obtain a proper estimate of the full error propagation in the unfolding procedure, we employ a toy-MC approach by sampling the deposited energy spectrum with Poisson fluctuations, and get the variations of the unfolded numbers of events in each incident energy bin. The root-mean-squares of the resulting B/C and B/O variations are taken as the $1\sigma$ statistical uncertainties.

The systematic uncertainties are investigated extensively in this analysis. Main sources of systematic uncertainties for the flux ratio measurements include the trigger efficiency, the charge selection, the background subtraction, the isotope composition of boron, the unfolding procedure, and the hadronic model.The HE trigger efficiency as a function of $E_{\rm BGO}$ is inferred by the data recorded with the low energy (LE) trigger, whose efficiency for boron to oxygen nuclei is almost 100\%. The HE trigger efficiencies of boron, carbon, and oxygen are measured to be higher than 95\%, 97\%, and 98\% for $E_{\rm BGO}>80$ GeV. The discrepancy of the HE trigger efficiency between MC simulations and the flight data is estimated to be within 2\% for boron, 0.5\% for carbon and oxygen. The resulting systematic uncertainty associated with the trigger efficiency for B/C and B/O is $\sim2.1\%$. 
For the measurement of flux ratios, most of the systematic uncertainties related to the event selection are cancelled out, except the charge selection. The efficiencies of STK charge selection and PSD charge selection are studied separately. The efficiency of the STK charge selection ($Q_{\rm STK1st}>4$) is estimated by a specific selected sample based on $Q_{\rm PSD}$ and the second STK layer charge $Q_{\rm STK2nd}$. The difference between MC simulations and the flight data is within 1\% for boron, 0.5\% for carbon and oxygen, resulting in a systematic uncertainty of 1.1\% for B/C and B/O. 
The efficiency of $Q_{\rm PSD}$ cut is estimated by a top-PSD layer based charge selection. The difference between MC simulations and the flight data is within 2\% for boron, 1\% for carbon and 2\% for oxygen. The corresponding systematic uncertainty is 2.2\% for B/C and 2.8\% for B/O.
The systematic uncertainty related to background estimate is investigated through varying the charge selection window of $Q_{\rm PSD}$ by $\pm0.1$ for boron, $\pm0.15$ for carbon and oxygen, and repeating the background estimate and the acceptance calculation. That results in energy-dependent variations of B/C and B/O, which are less than 2\% below 200 GeV/n and increases to $(6-7)\%$ at 5 TeV/n. 
The uncertainty due to the assumed boron isotope composition $Y_{\rm B}=0.7\pm0.1$ \cite{AMS:2016brs} is estimated to be 1.9\% for B/C and B/O, which is dominated by the calculation of the average atomic mass number of boron. The uncertainty related to spectral unfolding is evaluated via re-building the response matrix by varying the spectral index used to weight the MC simulations in a range of $[-3.3,-2.7]$ for boron, and $[-2.8,-2.4]$ for carbon and oxygen. The resulting variations of B/C and B/O are $\sim4\%$ for the first energy bin and less than 1\% for other higher energy bins. The systematic uncertainty due to hadronic interaction model is estimated by a comparison between different MC simulations, i.e. {\tt GEANT4} and {\tt FLUKA}. The difference is mainly due to the discrepancies on efficiency, background estimate and energy response. It varies from 1.7\% to 2.9\% for B/C and 2.8\% to 6.1\% for B/O.

The energy-dependent relative uncertainties for B/C and B/O are summarized in Fig.~\ref{fig-error}. The total systematic uncertainties are computed as the quadratic sum of all the components. We can see that the systematic uncertainties dominate over the statistical ones for energies below $\sim 1$ TeV/n and vice versa for high energies. The statistical uncertainties and the total systematic ones are presented separately in Table 1.

\subsection*{Spectral fitting}

In this work, we compare two models to describe the spectral features of the energy dependence 
of B/C and B/O. One is a power-law (PL) function
\begin{equation}
R(E_k)=R_0\left(\frac{E_k}{\rm GeV/n}\right)^{-\gamma},
\end{equation}
and the other is a broken power-law (BPL) function
\begin{equation}
R(E_k)=\left\{
\begin{aligned}
R_0\left(E_k/E_b\right)^{-\gamma_1}, ~~~ E_k\leq E_b\\
R_0\left(E_k/E_b\right)^{-\gamma_2}, ~~~ E_k> E_b
\end{aligned}\right.,
\end{equation}
where $E_b$ is the break energy. Note that the systematic uncertainties might be highly 
correlated, and we thus apply the nuisance parameter method \cite{2017PhRvD..95h2007A} 
to take such correlations into account. The $\chi^2$ function is 
\begin{equation}
\chi^2=\sum_{i}\sum_{j}[R(E_{k,i})S(E_{k,i};\,\boldsymbol{w})-R_i]{\mathcal C}^{-1}_{ij} [R(E_{k,j})S(E_{k,j};\,\boldsymbol{w})-R_j] + \sum_{\ell=1}^{m} \left(\frac{1-w_\ell}
{\tilde{\sigma}_{\rm sys,\ell}}\right)^2,
\end{equation}
where $E_{k,i}$, $R_i$, and $R(E_{k,i})$ are the median energy, measured ratio, and model
predicted ratio in the $i$-th energy bin respectively, ${\mathcal C}$ is the covariance 
matrix of the ratios derived from the toy MC simulation, $S(E_{i};\,\boldsymbol{w})$ is a 
piecewise function defined by its value $\boldsymbol{w}$, and $\tilde{\sigma}_{\rm sys,\ell}
=\sigma_{\rm sys}/R$ is the relative systematic uncertainty of the ratio in the energy
range covered by the $\ell$-th nuisance parameter. 
Note that the systematic uncertainty due to hadronic interaction models is singled out.
Here we consider 4 nuisance parameters, covering roughly half a decade in energy by each.

We first test the PL model. With the results based on the {\tt GEANT4} simulation, 
the best-fitting $\chi^2$ values are 42.35 for B/C and 57.81 for B/O, for the number of 
degree-of-freedom (dof) of 7. The fits are obviously poor.
The BPL model improves the fits significantly, resulting in $\chi^2/{\rm dof}=6.61/5$
and $5.51/5$ for B/C and B/O, respectively. We thus estimate that the measurements favor
the existing of breaks at $5.6\sigma$ and $6.9\sigma$ for the B/C and B/O ratios.
The fitting parameters of the BPL model are given in Table~\ref{tab:fit} and illustrated 
in Fig.~\ref{fig-fit}. The fits to the ratios based on the {\tt FLUKA} simulation give 
rather similar results, with a significance of the break at $4.4\sigma$ and $6.9\sigma$ for 
the B/C and B/O ratios, respectively. The differences due to these two simulation results 
are shown by the second error terms of Table~\ref{tab:fit}.
The break energies for the B/C fit and B/O fit are very close to each other,
$E_b \sim 100$ GeV/n. An additional uncertainty of the break energy associated with the 
energy measurement is about $2.2\%$ as estimated from the geomagnetic cutoff of electrons
and positrons \cite{Zang2018}.
The high-energy slopes of B/C and B/O above $E_b$ are consistent with each other, 
while the low-energy slope of B/C is slightly softer than that of B/O. We can also 
derive the slope changes below and above $E_b$, which is
$\Delta\gamma=0.155^{+0.026+0.000}_{-0.026-0.026}$ for B/C, and
$\Delta\gamma=0.207^{+0.027+0.000}_{-0.028-0.007}$ for B/O. 
Note for B/O, the systematic uncertainty on $\Delta\gamma$ due to hadronic interaction 
models is smaller than those for slopes $\gamma_1$ and $\gamma_2$, since both spectral indices
vary in the same direction.
It is shown that the B/O spectral hardening is slightly higher than the one present in B/C.
However, more precise measurements are necessary to confirm such a difference.

\begin{table}[ht]
\begin{center}
\caption{Parameters from the BPL fitting to the B/C and B/O ratios.} 
\begin{tabular}{lll} \hline\hline
& B/C  & B/O  \\
Nuisance parameters  & 4  &  4  \\ \hline
$R_0$ & $0.093^{+0.004+0.000}_{-0.004-0.001}$ & $0.084^{+0.003+0.000}_{-0.003-0.000}$ \\
$\gamma_1$  & $0.356^{+0.008+0.000}_{-0.008-0.017}$  & $0.394^{+0.010+0.000}_{-0.010-0.026}$ \\
$E_b$~(GeV/n) & $98.9^{+8.9+10.0}_{-8.8-0.0}$   & $99.5^{+7.4+7.7}_{-7.1-0.0}$ \\
$\gamma_2$ & $0.201^{+0.024+0.008}_{-0.024-0.000}$ & $0.187^{+0.024+0.000}_{-0.024-0.019}$ \\
$\chi^2$/dof & 6.61/5 & 5.51/5 \\ \hline\hline
\end{tabular}
\label{tab:fit}
\end{center}
\end{table}

\subsection*{Theoretical modelling}

Here we confront our measurements with predictions of several typical classes of models. 
Cowsik and Madziwa-Nussinov proposed a nested leaky box propagation model of CRs, with a constant residence time in the ISM and an energy-dependent residence time in cocoons surrounding the sources \cite{2016ApJ...827..119C}. This model predicts a constant B/C ratio at high energies with an increase at low energies, which represents a minimal extension of the leaky box model. In the more realistic diffusion model, the propagation of CRs in the Milky Way may also be inhomogeneous. Such a scenario was proposed to account for the hardenings of primary CRs, $\gamma$-ray observations, as well as large-scale anisotropies of CRs \cite{2012ApJ...752L..13T,Guo:2018wyf}. Zhao et al. performed a global fit to a comprehensive set of CR data to derive the parameters of the spatially-dependent propagation model \cite{Zhao:2021yzf}. The re-acceleration of CR particles by the random magnetic turbulence during the propagation can also result in a low-energy softening of the secondary-to-primary ratios \cite{2020JCAP...11..027Y}. In Fig.~\ref{fig-model}, we compare the predicted B/C ratios from the above models \cite{2016ApJ...827..119C,2020JCAP...11..027Y,Zhao:2021yzf} with the measurements of DAMPE. Also shown is the prediction with a break of the diffusion coefficient as discussed in the main text (red dashed line in Fig. 2. 
We note that most of the fittings were done with pre-DAMPE data, and a re-fitting including the DAMPE data may improve the goodness-of-fit. Nevertheless, the precise measurements of the B/C ratio by DAMPE can be useful in testing some of these models. For example, the constant high-energy ratio as predicted in the nested leaky box model seems to be less favored by our data. Also the re-acceleration effect may not be enough to produce a significant break of the B/C ratio. Critical tests of these models with more precise measurements of the secondary-to-primary CR ratios to even higher energies can be achieved in future.

\renewcommand\thefigure{S\arabic{figure}}
\setcounter{figure}{0}

\clearpage

\begin{figure}[!ht]
\begin{center}
\includegraphics[width=0.7\columnwidth]{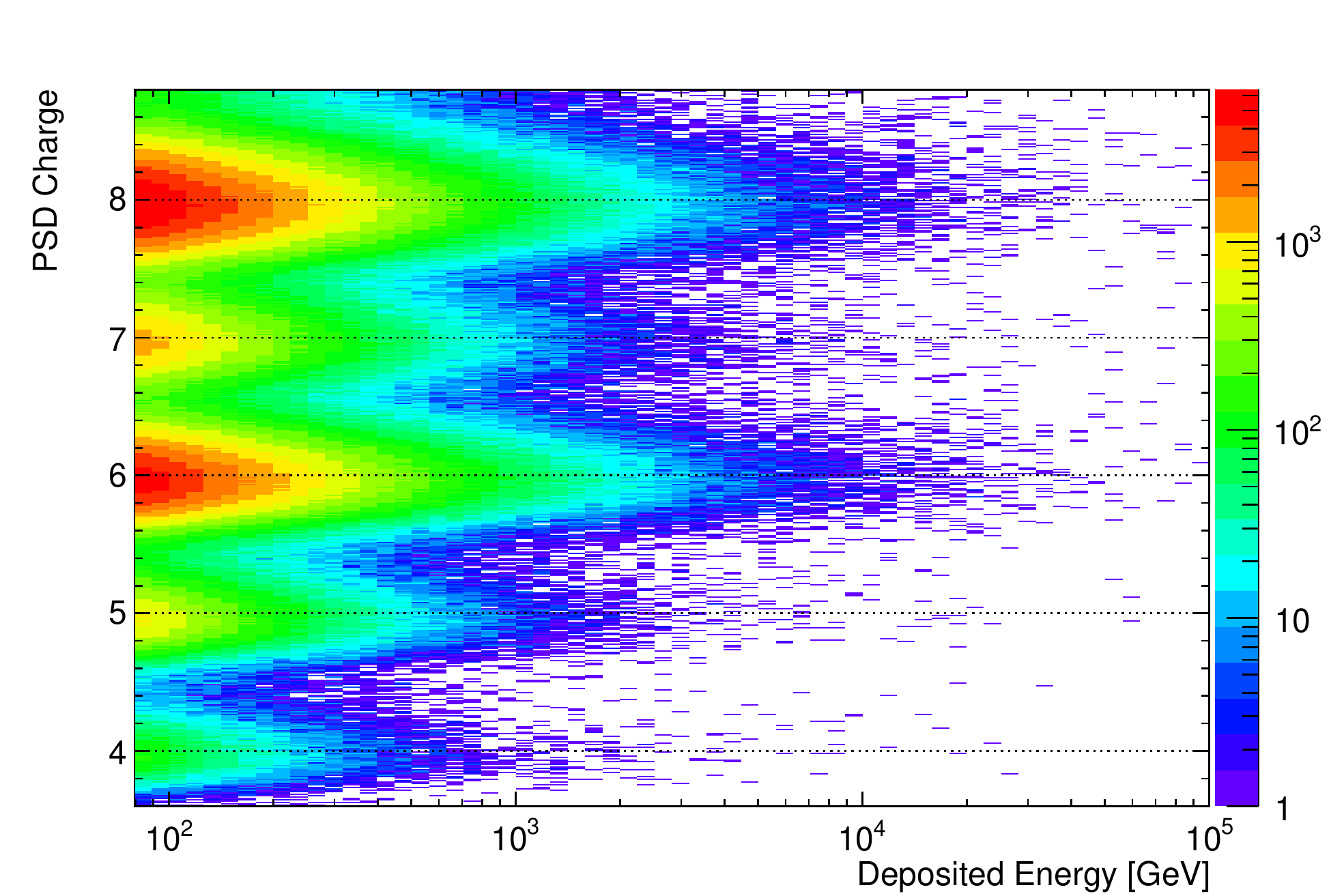}
\end{center}
\caption{
{\bf Particle charge reconstructed with PSD as a function of the deposited energy for particles with $Z=4-8$ in flight data.} 
}
\label{fig-psdQ}
\end{figure}

\begin{figure}[!ht]
\begin{center}
\includegraphics[width=0.48\columnwidth]{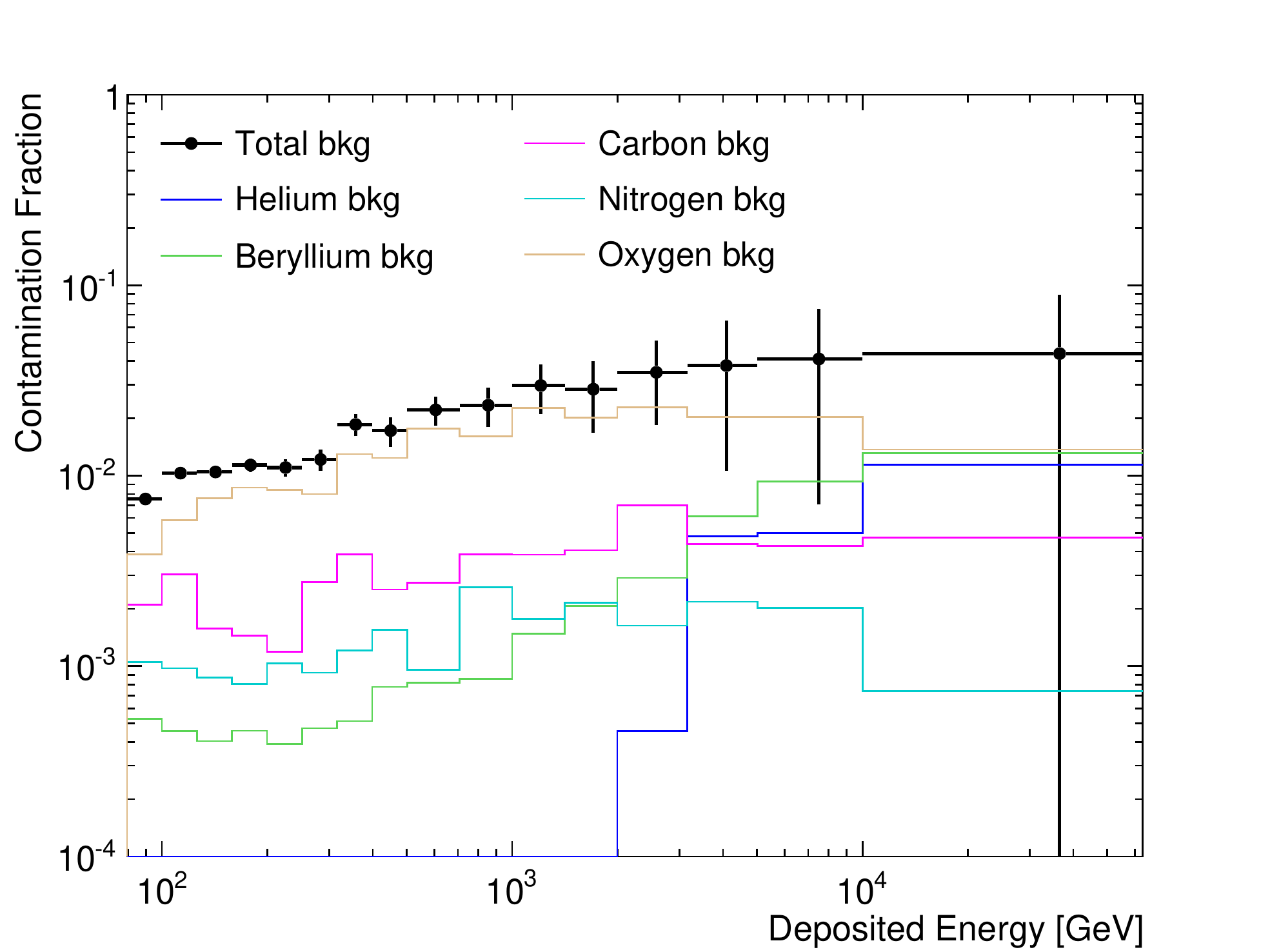}
\put(-35,135){\color{black}{\bf A}}
\includegraphics[width=0.48\columnwidth]{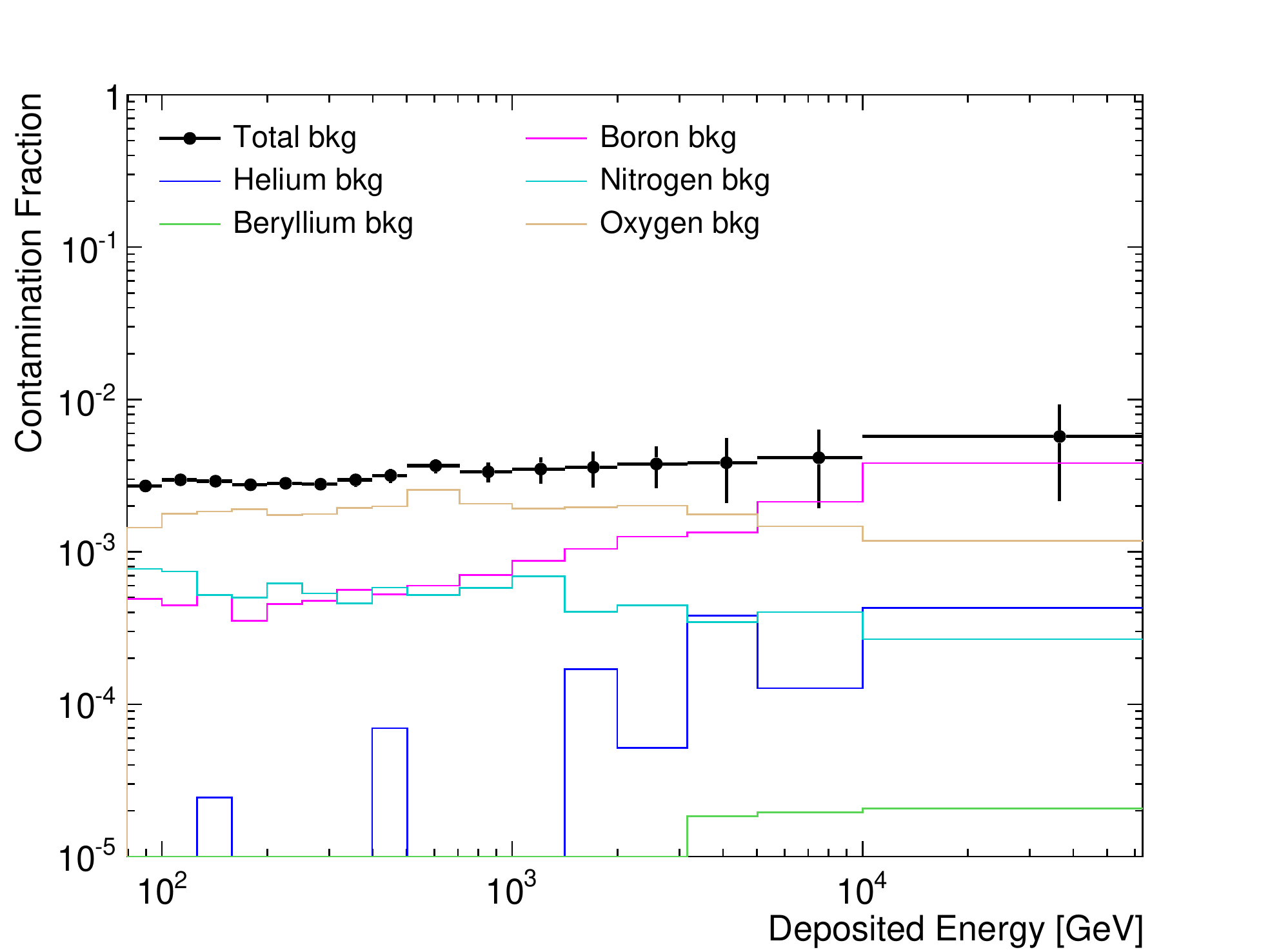}
\put(-35,135){\color{black}{\bf B}} \\
\includegraphics[width=0.48\columnwidth]{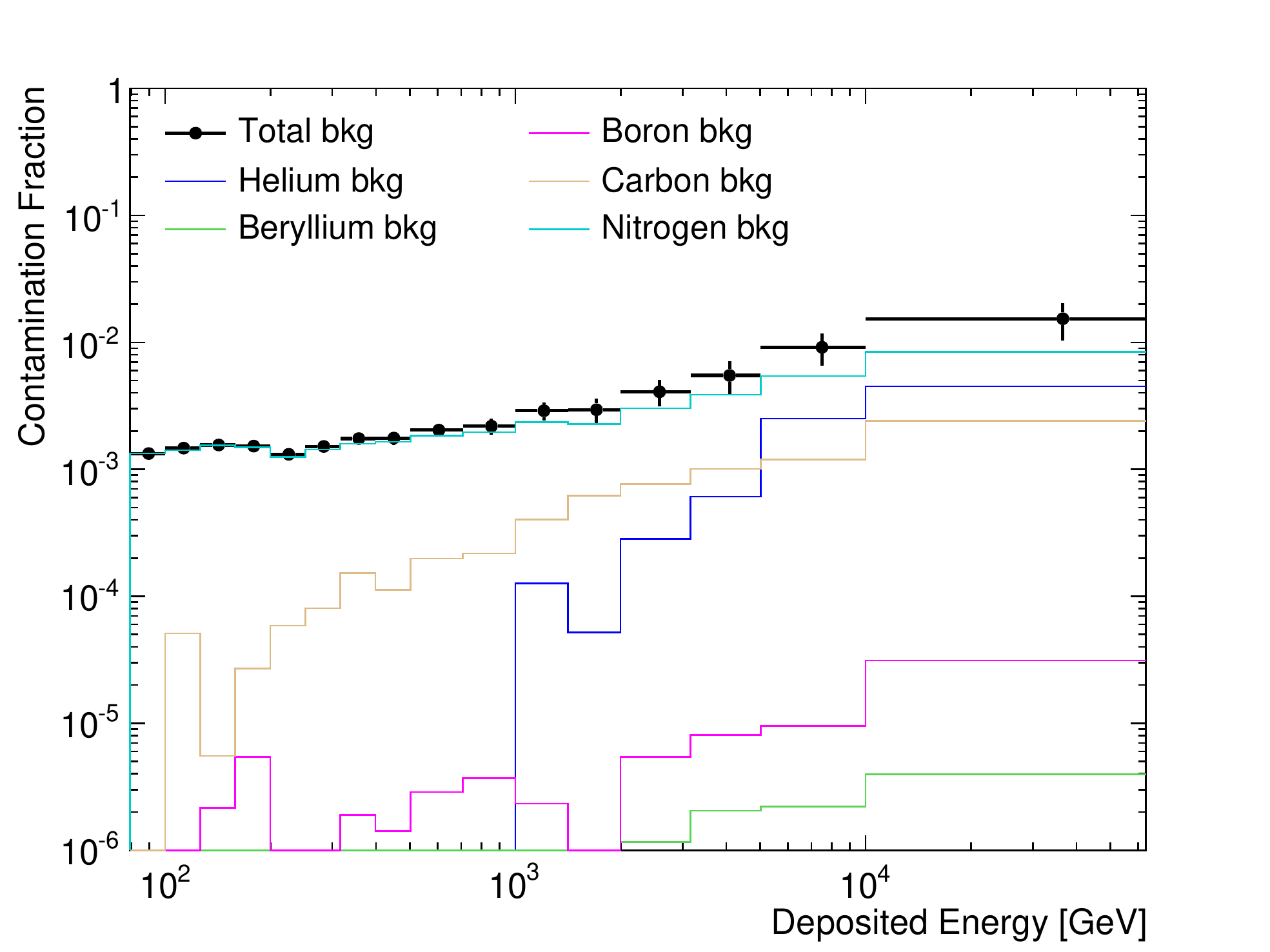}
\put(-35,135){\color{black}{\bf C}}
\end{center}
\caption{{\bf Fractions of background contamination from different particle species in selected sample of boron (A), carbon (B) and oxygen (C).} 
}
\label{fig-bkg}
\end{figure}

\begin{figure}[!ht]
\begin{center}
\includegraphics[width=0.48 \columnwidth]{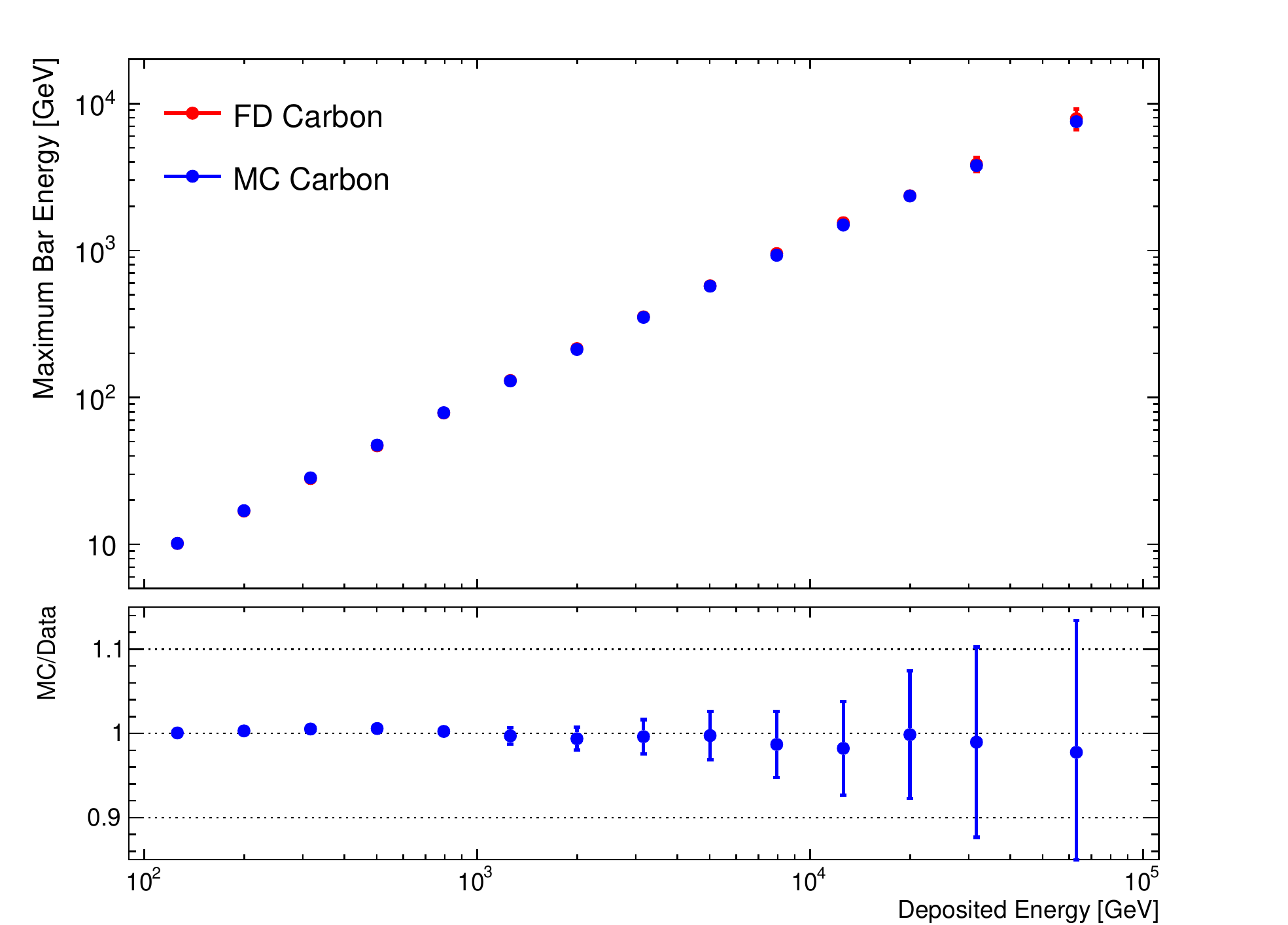}
\put(-40,70){\color{black}{\bf A}}
\includegraphics[width=0.48 \columnwidth]{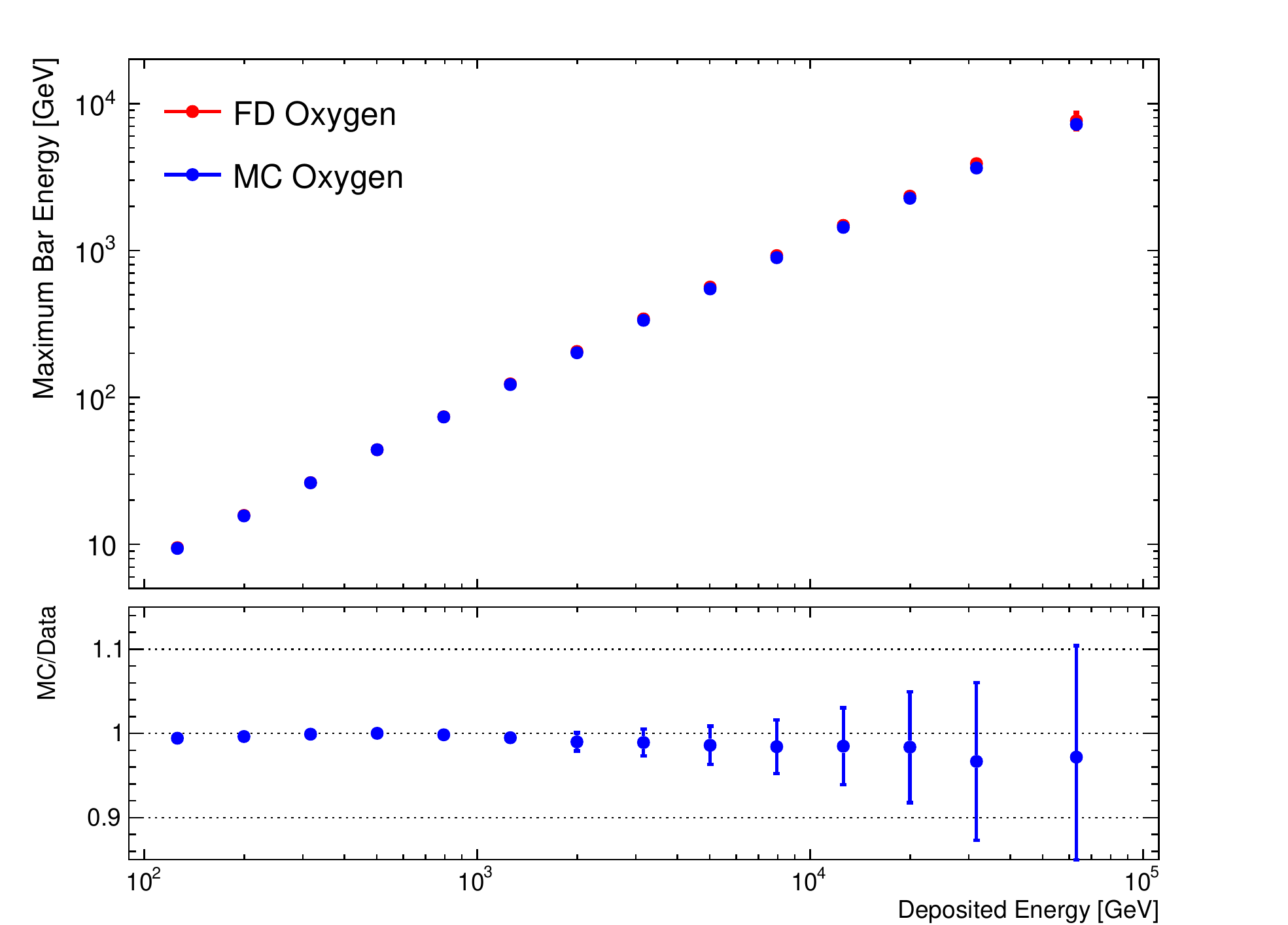}
\put(-40,70){\color{black}{\bf B}}
\end{center}
\caption{
{\bf Comparisons of the maximum energy deposited in one BGO crystal versus the total energy between flight data and {\tt GEANT4 FTFP\_BERT} simulations for carbon (A) and oxygen (B).}
}
\label{fig-maxbarE}
\end{figure}

\begin{figure}[!ht]
\begin{center}
\includegraphics[width=0.48 \columnwidth]{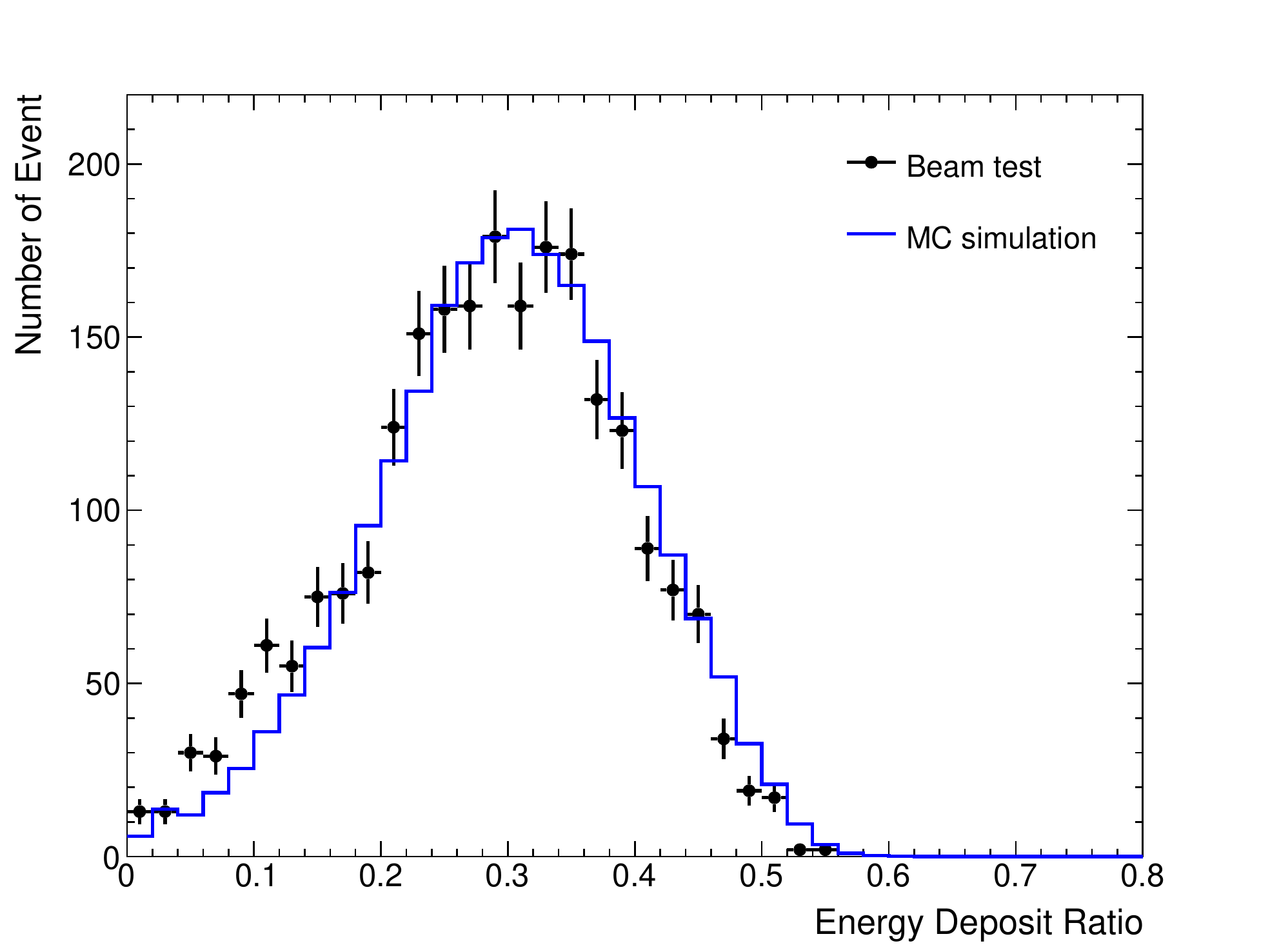}
\put(-190,135){\color{black}{\bf A}}
\includegraphics[width=0.48 \columnwidth]{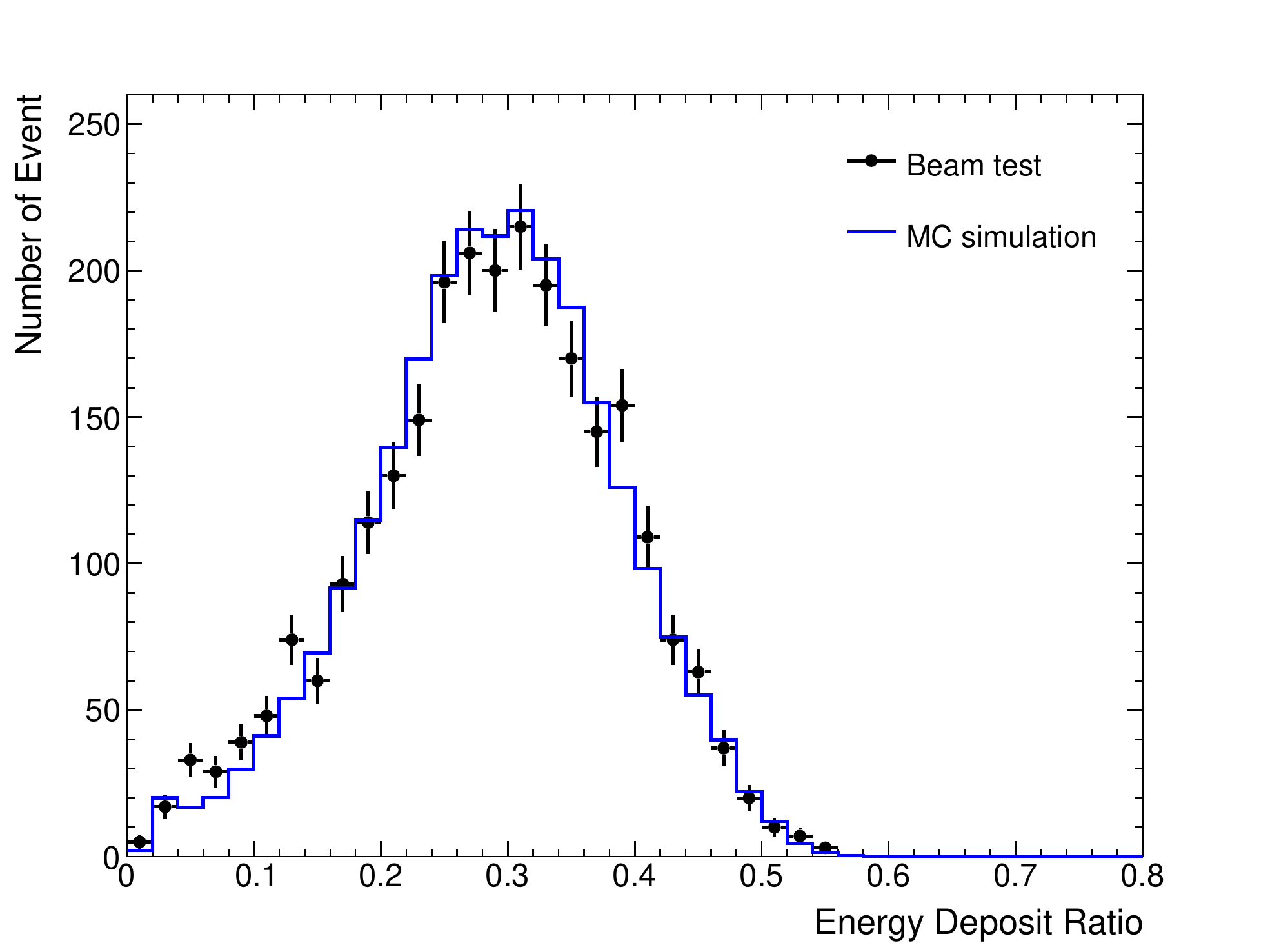}
\put(-190,135){\color{black}{\bf B}}
\end{center}
\caption{
{\bf Distributions of the fraction of deposited energy in the calorimeter for carbon (A) and oxygen (B) from test beams at CERN-SPS with kinetic energy of 75 GeV/n.}
The results from {\tt GEANT4 FTFP\_BERT} simulations (blue line) are overplotted for comparison.
}
\label{fig-beam}
\end{figure}

\begin{figure}[!ht]
\begin{center}
\includegraphics[width=0.48 \columnwidth]{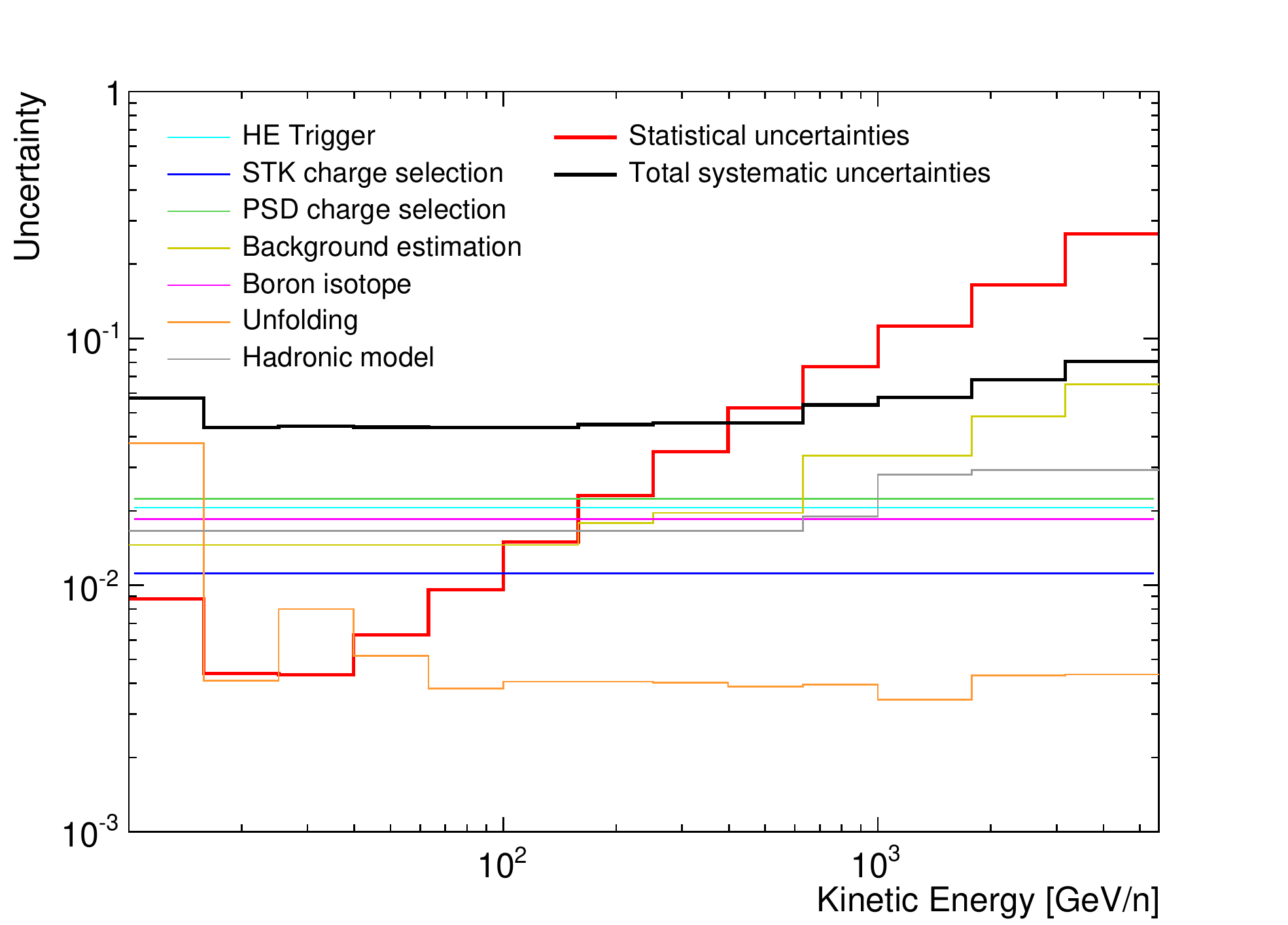}
\put(-35,130){\color{black}{\bf A}}
\includegraphics[width=0.48 \columnwidth]{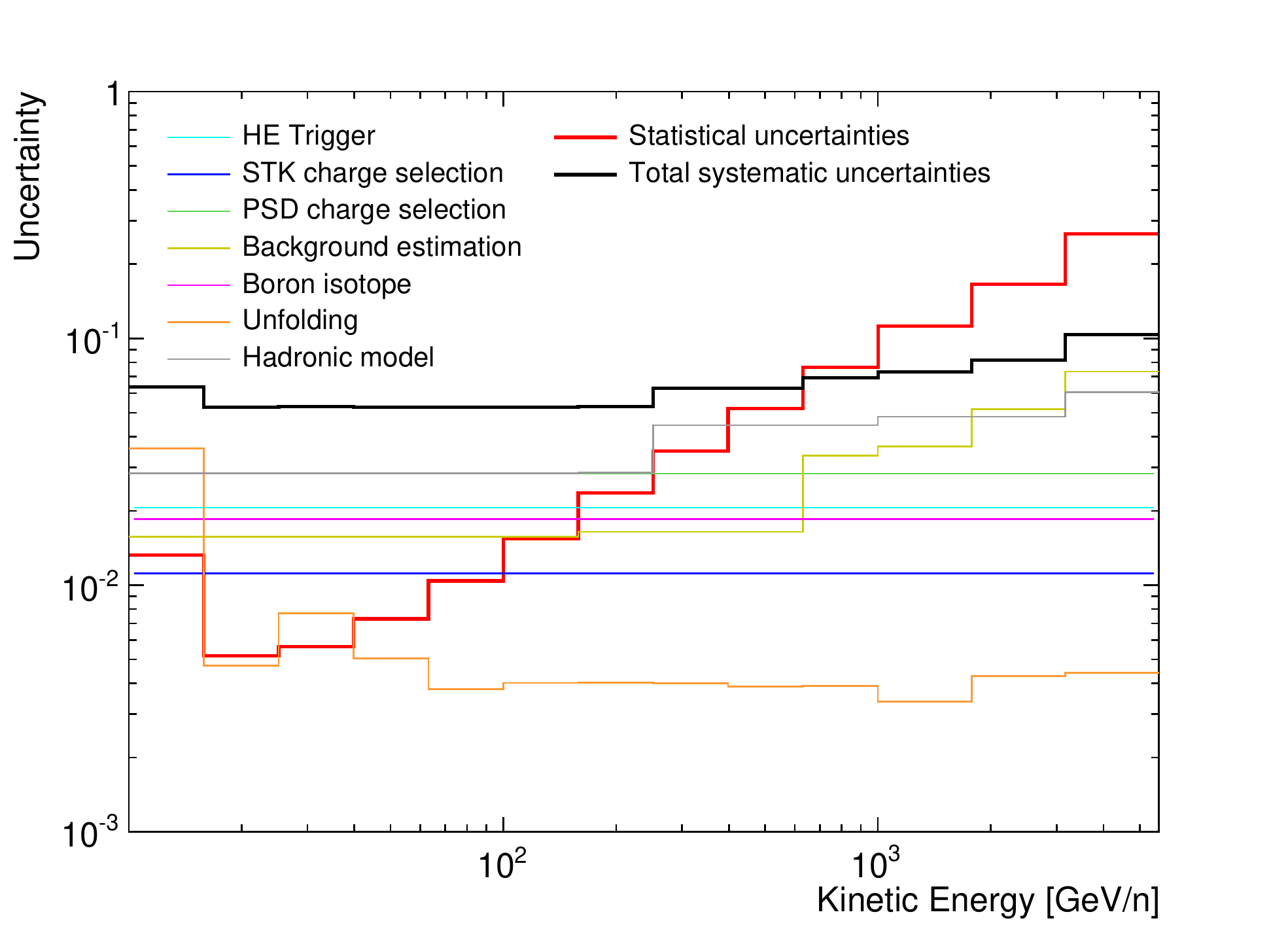}
\put(-35,130){\color{black}{\bf B}}
\end{center}
\caption{
{\bf Relative uncertainties of B/C (A) and B/O (B) as a function of the kinetic energy per nucleon.} 
The total systematic uncertainties are computed as the quadratic sum of all components.
}
\label{fig-error}
\end{figure}

\begin{figure}[!ht]
\begin{center}
\includegraphics[width=0.48 \columnwidth]{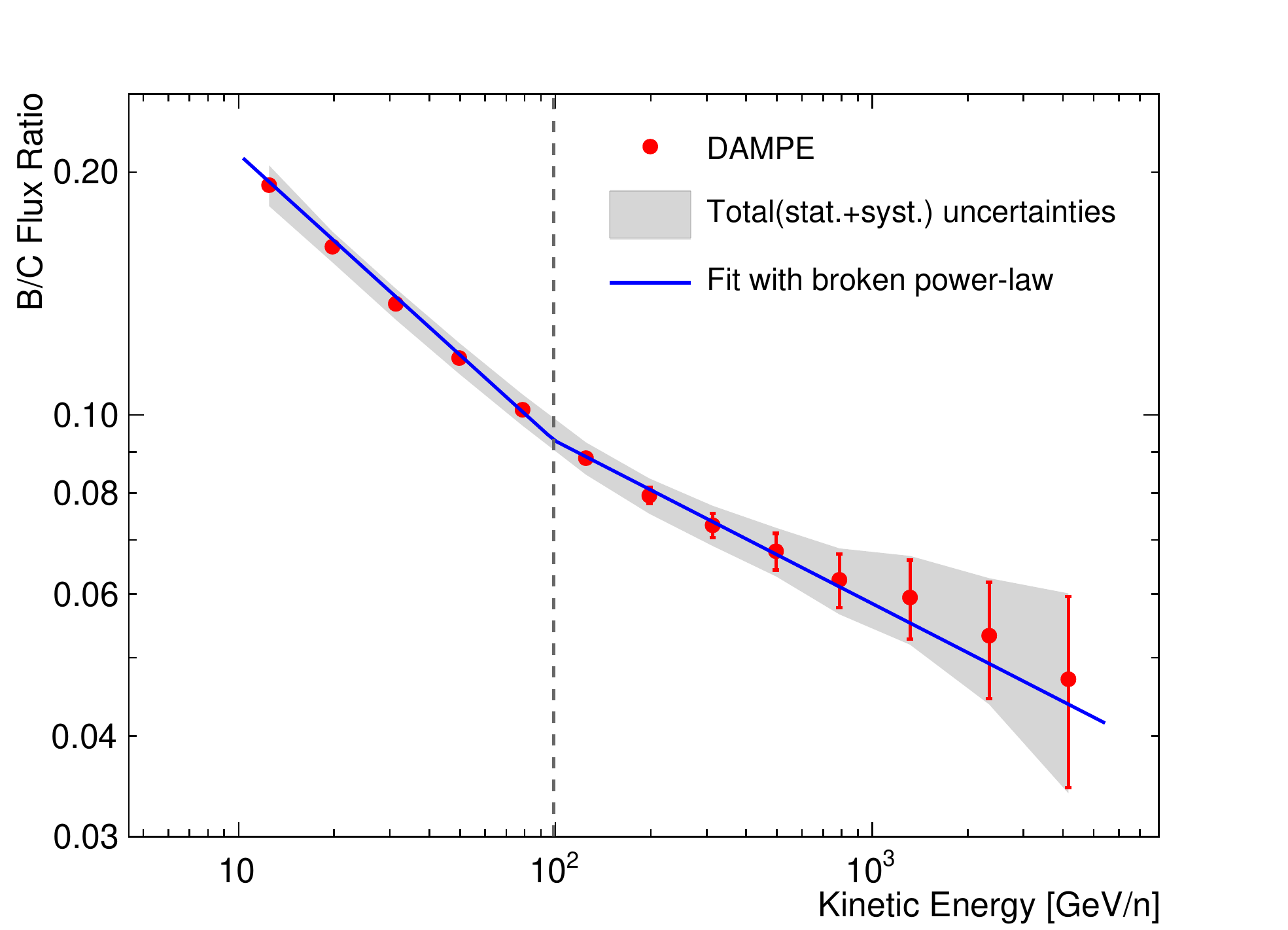}
\put(-190,25){\color{black}{\bf A}}
\includegraphics[width=0.48 \columnwidth]{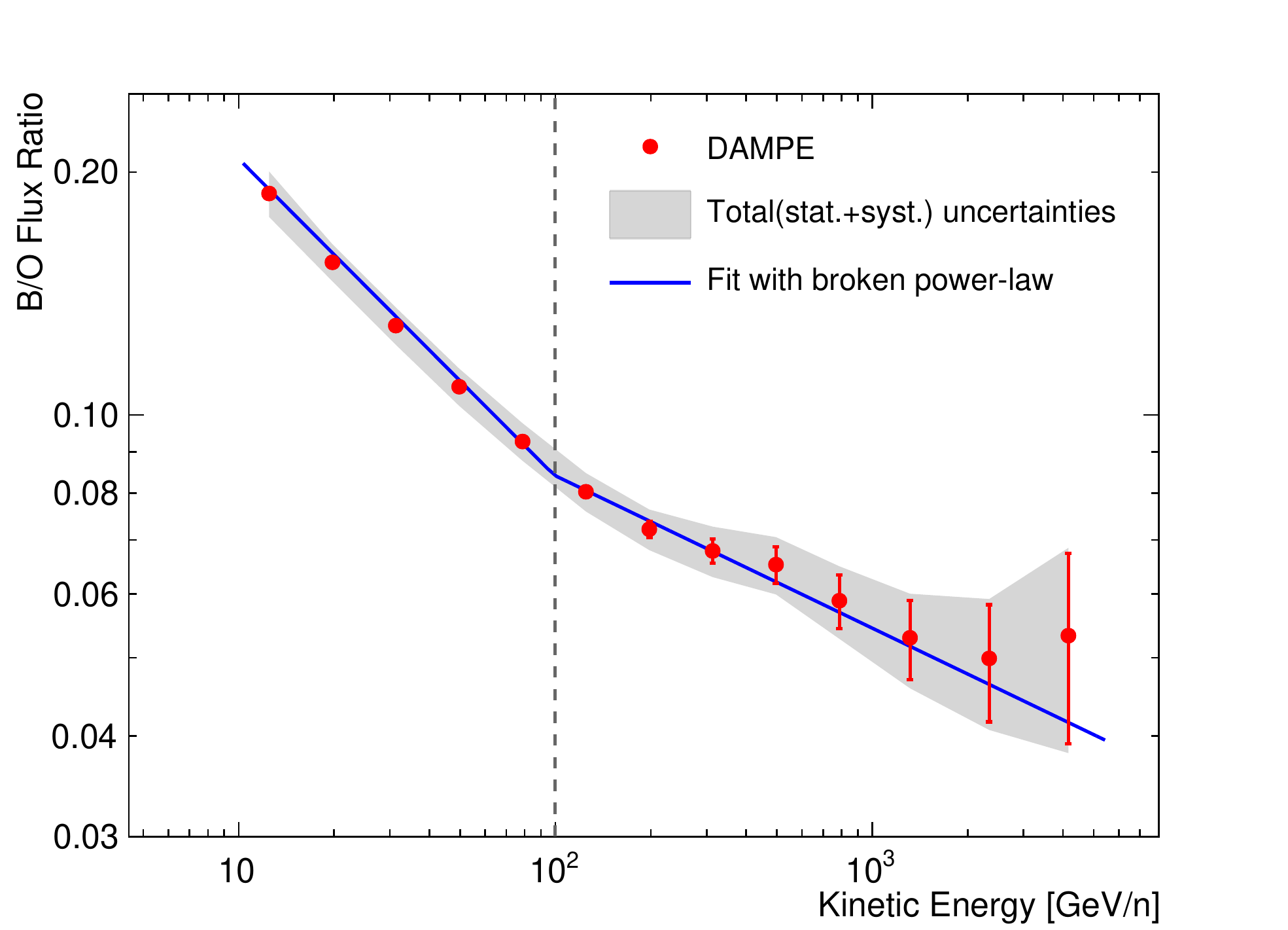}
\put(-190,25){\color{black}{\bf B}}
\end{center}
\caption{
{\bf 
Fits to the energy dependence of B/C (A) and B/O (B) with the BPL model.}
Gray dashed lines indicate the best fitted break energies.
} 
\label{fig-fit}
\end{figure}

\begin{figure}[!ht]
\begin{center}
\includegraphics[width=0.7\columnwidth]{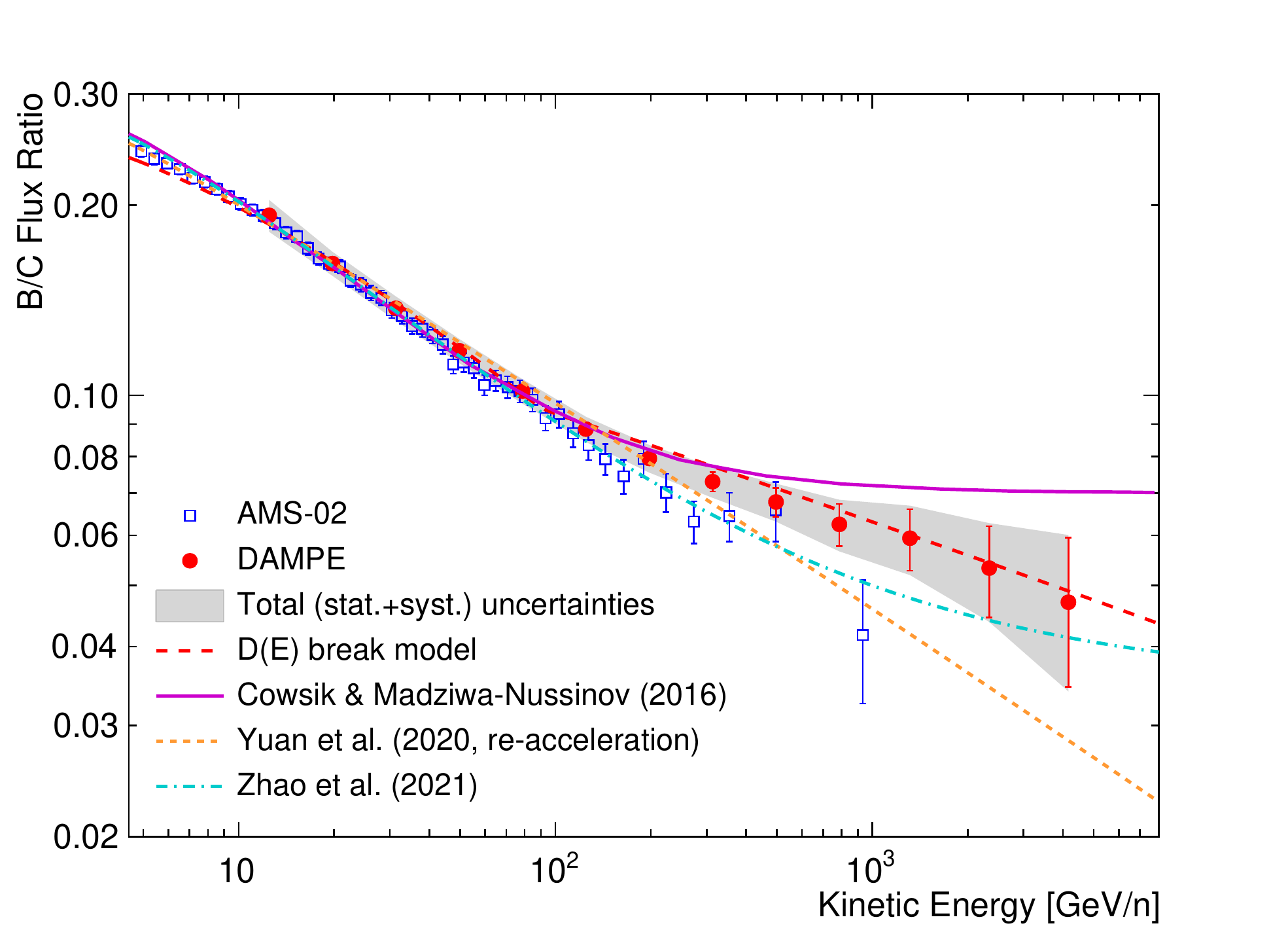}
\end{center}
\caption{
{\bf 
The predicted B/C ratios from different models, compared with the measurements from AMS-02 and DAMPE.}
} 
\label{fig-model}
\end{figure}

\clearpage

\end{document}